\documentclass[12pt]{article}
\usepackage{amsmath,amssymb,bm,epsfig,psfrag}

\allowdisplaybreaks 
\def\nb{\bar{n}}
\def\nslash{\rlap{\hspace{0.02cm}/}{n}}
\def\nbslash{\rlap{\hspace{0.02cm}/}{\bar n}}
\def\bT{\boldsymbol{T}}

\begin{document}

\begin{titlepage}

\begin{flushright}
FERMILAB-PUB-09-061-T\\
MZ-TH/09-05\\[2mm]
March 5, 2009 
\end{flushright}

\vspace{0.5cm}
\begin{center}
\Large\bf
On the Structure of Infrared Singularities of Gauge-Theory Amplitudes 
\end{center}

\vspace{0.5cm}
\begin{center}
{\sc Thomas Becher\,$^a$ and Matthias Neubert\,$^b$}\\
\vspace{0.4cm}
{\sl $^a$\,Fermi National Accelerator Laboratory,
P.O. Box 500, Batavia, IL 60510, U.S.A.\\
$^b$\,Institut f\"ur Physik (THEP), Johannes Gutenberg-Universit\"at,
D-55099 Mainz, Germany}
\end{center}

\vspace{0.2cm}
\begin{abstract}\noindent
A closed formula is obtained for the infrared singularities of dimensionally regularized, massless gauge-theory scattering amplitudes with an arbitrary number of legs and loops. It follows from an all-order conjecture for the anomalous-dimension matrix of $n$-jet operators in soft-collinear effective theory. We show that the form of this anomalous dimension is severely constrained by soft-collinear factorization, non-abelian exponentiation, and the behavior of amplitudes in collinear limits. Using a diagrammatic analysis, we demonstrate that these constraints imply that to three-loop order the anomalous dimension involves only two-parton correlations, with the possible exception of a single color structure multiplying a function of conformal cross ratios depending on the momenta of four external partons, which would have to vanish in all two-particle collinear limits. We suggest that such a function does not appear at three-loop order, and that the same is true in higher orders. Our formula predicts Casimir scaling of the cusp anomalous dimension to all orders in perturbation theory, and we explicitly check that the constraints exclude the appearance of higher Casimir invariants at four loops. Using known results for the quark and gluon form factors, we derive the three-loop coefficients of the $1/\epsilon^n$ pole terms (with $n=1,\dots,6$) for an arbitrary $n$-parton scattering amplitude in massless QCD. This generalizes Catani's two-loop formula proposed in 1998. 
\end{abstract}
\vfil

\end{titlepage}

\tableofcontents
\newpage

\section{Introduction}

The origin and structure of ultraviolet (UV) divergences in quantum field theories is well understood. They can be absorbed into a renormalization of the parameters of the theory. The fact that physical results must be independent of the UV regulator introduced in intermediate steps of a calculation gives rise to powerful constraints, which are summarized by the renormalization-group (RG) equations of the theory. Perturbative results for on-shell scattering amplitudes in theories with massless fields also contain infrared (IR) singularities, which originate from loop-momentum configurations where particle momenta become soft or collinear. These singularities cancel in physical observables, which also include real radiation and are insensitive to soft and collinear emissions \cite{Kinoshita:1962ur,Lee:1964is}. 

In a recent letter \cite{Becher:2009cu}, we have shown that the IR singularities of on-shell, $n$-particle scattering amplitudes in massless QCD are in one-to-one correspondence with UV divergences of operators defined in soft-collinear effective theory (SCET) \cite{Bauer:2000yr,Bauer:2001yt,Bauer:2002nz,Beneke:2002ph}. This implies that they can be analyzed using standard methods of operator renormalization. In particular, the IR divergences of $n$-point scattering amplitudes can be absorbed into a multiplicative renormalization factor $\bm{Z}$ and are fully determined by an anomalous dimension $\bm{\Gamma}$. Both the $\bm{Z}$-factor and the anomalous dimension are matrices in color space, i.e., they mix amplitudes with the same particle content but different color structure. In dimensional regularization, IR singularities manifest themselves as poles in $\epsilon=(2-d/2)$. The renormalization conditions imply that the higher $1/\epsilon^n$ poles are determined by the single-pole coefficient, from which one obtains the anomalous dimension. In contrast to traditional effective theories, SCET operators are nonlocal along the direction of large momentum flow. This induces a dependence of the anomalous dimension $\bm{\Gamma}$ on the momentum transfers and translates into the presence of Sudakov double logarithms in high-energy scattering processes.

The structure of the effective theory imposes non-trivial constraints on the form of the anomalous dimension $\bm{\Gamma}$. SCET contains a separate set of collinear fields for each direction of large energy flow, and a single set of soft fields mediating interactions among the different collinear sectors. Up to contributions suppressed by powers of the large momentum transfers, the different collinear fields interact only through soft gluon exchange. The matrix elements of $n$-parton operators factor into jet functions associated with each set of collinear fields and a soft function which arises from interactions among the different jets. Since different collinear fields do not interact, the jet functions are color diagonal. Non-trivial momentum dependence and color correlations of the anomalous dimension arise from soft interactions. The fact that the soft and jet functions must conspire to produce an anomalous dimension depending only on hard momentum transfers leads to strong constraints on the form of the allowed terms.

Due to the eikonal form of soft interactions, they can be represented as Wilson lines along the directions of the external colored particles. This leads to further constraints implied by the non-abelian exponentiation theorem \cite{Gatheral:1983cz,Frenkel:1984pz}. In QED, eikonal identities imply that higher-order soft radiation is obtained by exponentiating the leading-order contribution \cite{Yennie:1961ad}. In QCD this is no longer true, but only a particular set of color structures can contribute to the soft anomalous dimension. We argue that their coefficients are related to each other by the structure of eikonal interactions. Additional constraints are obtained by considering limits in which some of the partons involved in a scattering process become collinear. In the simplest case where two external momenta become collinear, an $n$-parton amplitude reduces to a sum of $(n-1)$-parton amplitudes multiplied by splitting amplitudes. These splitting amplitudes and their anomalous dimensions can only depend on the momenta and quantum numbers of the particles involved in the splitting process. This leads to a constraint on the difference of the anomalous dimensions of the amplitudes before and after the splitting.

In this paper we derive the most general form of the anomalous dimension $\bm{\Gamma}$ compatible with the above constraints. We argue that to all orders in perturbation theory the anomalous dimension has the same structure as at one-loop order, featuring only two-particle correlations in colors and momenta. Our result is semi-classical in that it describes color-dipole interactions, which probe the momenta and color charges of pairs of external particles. This is reminiscent of Low's theorem, which relates the cross section for soft photon emission to the classical cross section times a factor depending on the electric charges and momenta of the external particles \cite{Low:1958sn}. In our case, the quantum structure of the underlying field theory shows itself via the coefficient of the color-dipole operator, which we relate to the universal cusp anomalous dimension of light-like Wilson loops, and via anomalous dimensions for quark and gluon fields. In practice, these anomalous dimensions can be determined at three-loop order using existing results for the quark and gluon form factors in QCD. However, our analysis is completely general and extends to arbitrary gauge theories with massless fields.

Starting at three-loop order, the analysis of the structure of the anomalous dimension becomes non-trivial. Based on a general study of the constraints following from soft-collinear factorization and the non-abelian exponentiation theorem, we show that in the most general case two additional color structures beyond those predicted by our simple formula could appear at three-loop order. One of them describes interactions among four different partons. We also show that at four-loop order a new structure could arise, which would violate Casimir scaling of the cusp anomalous dimension. However, imposing the correct behavior of $n$-parton scattering amplitudes in the limit where two partons become collinear eliminates all of these additional structures. We also present a second argument for the absence of the additional terms, based on the simple form of color-symmetrized soft-gluon interactions, which suggests that these new structures vanish due to color conservation. In this way our formula is established at three-loop order. We note, however, that our arguments could be circumvented by functions of conformal cross ratios of four parton momenta that vanish whenever two partons become collinear. At three-loop order a single color structure involving such a function is allowed by non-abelian exponentiation. For our conjecture to be valid, this function must vanish. The observation of Casimir scaling of the four-loop cusp anomalous dimension is unaffected by this caveat.

The arguments presented in this work suggest that our conjecture may
hold to all orders in perturbation theory. This would imply a new set of exact relations among amplitudes in perturbative quantum field theory and will hopefully shed new light on their deeper structure. The knowledge of the structure of IR singularities, and of the associated anomalous-dimension matrix, is also of significant practical interest for collider physics. It is a necessary ingredient for the resummation of large logarithms in jet-production processes. Solving RG equations in SCET resums perturbative logarithms of ratios of the invariant masses of the produced jets to the large momentum transfers involved in their production. As an application of our results, we predict the complete structure of IR singularities for $n$-parton scattering amplitudes in massless QCD at the three-loop order. In recent work, the four-gluon amplitude in ${\cal N}=4$ super-Yang-Mills theory (SYM) was studied at three-loop order and expressed in terms of a small set of master integrals \cite{Bern:2008pv}. Once these integrals have been evaluated analytically or in numerical form, this calculation will provide a stringent test of our predictions. Another interesting implication of our analysis is the prediction that the cusp anomalous dimensions of quarks and gluons should obey Casimir scaling to all orders in perturbation theory, i.e., they should be equal to the quadratic Casimir operator $C_R$ in the fundamental or adjoint representation times a universal coefficient. While Casimir scaling has been shown to hold up to three-loop order by explicit calculation \cite{Moch:2004pa}, this prediction is highly non-trivial in view of the expectation that this scaling should no longer hold non-perturbatively, at least not for the finite parts of Wilson-loop expectation values. Already a long time ago, Frenkel and Taylor argued that Casimir scaling would be inconsistent with expectations about the area law for matrix elements of Wilson loops giving rise to confinement \cite{Frenkel:1984pz}. More recently, investigations of high-spin operators in string theory using the AdS/CFT correspondence \cite{Maldacena:1997re,Gubser:1998bc,Witten:1998qj} have found a strong-coupling behavior that is inconsistent with Casimir scaling \cite{Armoni:2006ux,Alday:2007hr,Alday:2007mf}. 

We begin in Section~\ref{sec:RG} by recalling the connection between on-shell scattering amplitudes and Wilson coefficient functions in SCET \cite{Becher:2009cu}. We then solve the RG equation for the renormalization factor $\bm{Z}$ in terms of an integral over the anomalous-dimension matrix of $n$-jet SCET operators. We derive the three-loop expression for $\bm{Z}$ in terms of known anomalous-dimension coefficients and show that our approach reproduces Catani's result for the divergences of two-loop amplitudes \cite{Catani:1998bh}. In Section~\ref{sec:SCET} we discuss the structure of SCET for $n$-jet processes, explicitly construct the necessary operators and show that their matrix elements factor into jet and soft functions. We stress the importance of soft operators built out of $n$ light-like Wilson lines, which describe the color and momentum correlations in the anomalous-dimension matrix. A detailed discussion of the arguments supporting our conjecture for the structure of the anomalous-dimension matrix is presented in Section~\ref{sec:arguments}. We recall important facts about the renormalization of Wilson loops with cusps and cross points, the non-abelian exponentiation theorem, and soft-collinear factorization in SCET. In Section~\ref{sec:collinear} we study the implications of the known behavior of scattering amplitudes in the limit where two of the external partons become collinear. We find that the $n$-jet anomalous-dimension matrix can be decomposed in this limit into the sum of an $(n-1)$-jet anomalous dimension and the anomalous dimension of the splitting amplitudes, whose all-order form we derive. The nature of this decomposition imposes another strong constraint on the momentum and color structures that can appear in the anomalous dimension. In Section~\ref{sec:diagrammar} we perform an explicit analysis of the structure of the soft anomalous-dimension matrix to three-loop order. We first list all structures allowed by non-abelian exponentiation and then impose the constraints from soft-collinear factorization and from two-particle collinear limits. These constraints eliminate all additional terms with the exception of a single color structure, which is of subleading order in the $N_c\to\infty$ limit and compatible with the constraints if it is multiplied by a function of conformal ratios which vanishes in all collinear limits. We also show that the constraints enforce Casimir scaling of the cusp anomalous dimension to four loops. Our conclusions are presented in Section~\ref{sec:summary}. Perturbative results required to evaluate our formulae at three-loop order and some comments on the behavior of our color structures in the large-$N_c$ limit are compiled in two appendices.

\section{IR factorization and RG invariance}
\label{sec:RG}

The key observation of our letter \cite{Becher:2009cu} was that the IR singularities of on-shell amplitudes in massless QCD are in one-to-one correspondence to the UV poles of operator matrix elements in SCET. These poles can therefore be subtracted by means of a multiplicative renormalization factor $\bm{Z}$, which is a matrix in color space. Specifically, we have shown that the finite remainders of the scattering amplitudes can be obtained from the IR divergent, dimensionally regularized amplitudes via the relation
\begin{equation}\label{renorm}
   |{\cal M}_n(\{\underline{p}\},\mu)\rangle 
   = \lim_{\epsilon\to 0}\,
   \bm{Z}^{-1}(\epsilon,\{\underline{p}\},\mu)\,
   |{\cal M}_n(\epsilon,\{\underline{p}\})\rangle \,.
\end{equation}
Here $\{\underline{p}\}\equiv\{p_1,\dots,p_n\}$ represents the set of the momentum vectors of the $n$ partons, and $\mu$ denotes the factorization scale. The quantity $|{\cal M}_n(\epsilon,\{\underline{p}\})\rangle$ on the right-hand side is a UV-renormalized, on-shell $n$-parton scattering amplitude with IR singularities regularized in $d=4-2\epsilon$ dimensions. After coupling constant renormalization, these amplitudes are UV finite. Apart from trivial spinor factors and polarization vectors for the external particles, the minimally subtracted scattering amplitudes $|{\cal M}_n(\{\underline{p}\},\mu)\rangle$ on the left-hand side of (\ref{renorm}) coincide with Wilson coefficients of $n$-jet operators in SCET \cite{Becher:2009cu}, to be defined later:
\begin{equation}\label{MnCn}
   |{\cal M}_n(\{\underline{p}\},\mu)\rangle 
   = |{\cal C}_n(\{\underline{p}\},\mu)\rangle 
   \times \mbox{[on-shell spinors and polarization vectors]\,.} 
\end{equation}
We postpone a more detailed discussion of the effective theory to Section~\ref{sec:SCET} and proceed to study the implications of this observation. 

To analyze the general case of an arbitrary $n$-parton amplitude, it is convenient to use the color-space formalism of \cite{Catani:1996jh,Catani:1996vz}, in which amplitudes are treated as $n$-dimensional vectors in color space. $\bm{T}_i$ is the color generator associated with the $i$-th parton in the scattering amplitude, which acts as an $SU(N_c)$ matrix on the color indices of that parton. Specifically, one assigns $(\bm{T}_i^a)_{\alpha\beta}=t_{\alpha\beta}^a$ for a final-state quark or initial-state anti-quark, $(\bm{T}_i^a)_{\alpha\beta}=-t_{\beta\alpha}^a$ for a final-state anti-quark or initial-state quark, and $(\bm{T}_i^a)_{bc}=-if^{abc}$ for a gluon. We also use the notation $\bm{T}_i\cdot\bm{T}_j\equiv \bm{T}_i^a\,\bm{T}_j^a$ summed over $a$. Generators associated with different particles trivially commute, $\bm{T}_i\cdot\bm{T}_j=\bm{T}_j\cdot\bm{T}_i$ for $i\ne j$, while $\bm{T}_i^2=C_i$ is given in terms of the quadratic Casimir operator of the corresponding color representation, i.e., $C_q=C_{\bar q}=C_F$ for quarks or anti-quarks and $C_g=C_A$ for gluons. Because they conserve color, the scattering amplitudes fulfill the relation 
\begin{equation}\label{singlet}
   \sum_i\,\bm{T}_i^a\,
   |{\cal M}_n(\epsilon,\{\underline{p}\})\rangle = 0 \,.
\end{equation}

It follows from (\ref{renorm}) that the minimally subtracted scattering amplitudes satisfy the RG equation
\begin{equation}\label{RGE}
   \frac{d}{d\ln\mu}\,|{\cal M}_n(\{\underline{p}\},\mu)\rangle
   = \bm{\Gamma}(\{\underline{p}\},\mu)\,
   |{\cal M}_n(\{\underline{p}\},\mu)\rangle \,,
\end{equation}
where the anomalous dimension is related to the $\bm{Z}$-factor by 
\begin{equation}\label{Gammadef}
   \bm{\Gamma}(\{\underline{p}\},\mu)
   = - \bm{Z}^{-1}(\epsilon,\{\underline{p}\},\mu)\,
   \frac{d}{d\ln\mu}\,\bm{Z}(\epsilon,\{\underline{p}\},\mu) \,.
\end{equation}
The formal solution to this equation can be written in the form
\begin{equation}\label{RGEsol}
   \bm{Z}(\epsilon,\{\underline{p}\},\mu) 
   = {\rm\bf{P}} \exp\left[ \int_\mu^{\infty} 
   \frac{{\rm d}\mu'}{\mu'}\,\bm{\Gamma}(\{\underline{p}\},\mu') 
   \right] ,
\end{equation}
where the path-ordering symbol $\rm\bf{P}$ means that matrices are ordered from left to right according to decreasing values of $\mu'$. The upper integration value follows from asymptotic freedom and the fact that 
$\bm{Z}=\bm{1}+{\cal O}(\alpha_s)$.

In Section~\ref{sec:arguments}, we will discuss theoretical arguments supporting an all-order conjecture for the anomalous-dimension matrix presented in \cite{Becher:2009cu}, which states that it has the simple form
\begin{equation}\label{magic}
   \bm{\Gamma}(\{\underline{p}\},\mu) 
   = \sum_{(i,j)}\,\frac{\bm{T}_i\cdot\bm{T}_j}{2}\,
    \gamma_{\rm cusp}(\alpha_s)\,\ln\frac{\mu^2}{-s_{ij}} 
    + \sum_i\,\gamma^i(\alpha_s) \,,
\end{equation}
where $s_{ij}\equiv 2\sigma_{ij}\,p_i\cdot p_j+i0$, and the sign factor $\sigma_{ij}=+1$ if the momenta $p_i$ and $p_j$ are both incoming or outgoing, and $\sigma_{ij}=-1$ otherwise. Here and below the sums run over the $n$ external partons. The notation $(i_1,...,i_k)$ refers to unordered tuples of distinct parton indices. Our result features only pairwise correlations among the color charges and momenta of different partons. These are the familiar color-dipole correlations arising already at one-loop order from a single soft gluon exchange. The fact that higher-order quantum effects do not induce more complicated structures and multi-particle correlations indicates a semi-classical origin of IR singularities. Besides wave-function-renormalization-type subtractions accomplished by the single-particle terms $\gamma^i$, the only quantum aspect appearing in (\ref{magic}) is a universal anomalous-dimension function $\gamma_{\rm cusp}$ related to the cusp anomalous dimension of Wilson loops with light-like segments \cite{Korchemsky:1987wg,Korchemsky:1988hd,Korchemskaya:1992je}. The three anomalous-dimension functions entering our result are defined by relation (\ref{magic}). They can be extracted from the known IR divergences of the on-shell quark and gluon form factors, which have been calculated to three-loop order \cite{Moch:2005id,Moch:2005tm,Baikov:2009bg}. The explicit three-loop expressions are given in Appendix~A.

Concerning the form of (\ref{magic}), we note that a conjecture that an analogous expression for the soft anomalous-dimension matrix (see Section~\ref{sec:45} below) might hold to all orders was mentioned in passing in the introduction of \cite{Bern:2008pv}, without presenting any supporting arguments. In a very recent paper, Gardi and Magnea have analyzed the soft anomalous-dimension matrix in more detail and found that (\ref{magic}) is the simplest solution to a set of constraints they have derived \cite{Gardi:2009qi}. However, they concluded that the most general solution could be considerably more complicated. Indeed, we emphasize that as a consequence of our result some amazing cancellations must occur in multi-loop calculations of scattering amplitudes. At $L$-loop order Feynman diagrams can involve up to $2L$ parton legs, while the most non-trivial graphs without subdivergences can still connect $(L+1)$ partons. We predict that these complicated diagrams can be decomposed into two-particle terms, whose color and momentum structures resemble that of one-loop diagrams. At two-loop order, these cancellations were found by explicit calculation in \cite{MertAybat:2006wq,MertAybat:2006mz}. More recently, the analysis was extended to the subclass of three-loop graphs containing fermion loops \cite{Dixon:2009gx}. In Section~\ref{sec:2loop} we will present a simple symmetry argument explaining these results.

To derive the perturbative expansion of the $\bm{Z}$-factor from the formal solution (\ref{RGEsol}) we use the generalized expression 
\begin{equation}
   \frac{d\alpha_s}{d\ln\mu} = \beta(\alpha_s,\epsilon)
   = \beta(\alpha_s) - 2\epsilon\,\alpha_s
\end{equation} 
for the $\beta$-function in $d=4-2\epsilon$ dimensions, where $\alpha_s\equiv\alpha_s(\mu)$ is the renormalized coupling constant. The simple form of (\ref{magic}) implies that the matrix structure of the anomalous dimension is the same at all scales, i.e., $[\bm{\Gamma}(\{\underline{p}\},\mu_1),\bm{\Gamma}(\{\underline{p}\},\mu_2)]=0$. The path-ordering symbol can thus be dropped in (\ref{RGEsol}), and we can directly obtain an expression for the logarithm of the renormalization factor. Writing $\bm{\Gamma}(\{\underline{p}\},\mu,\alpha_s(\mu))$ instead of $\bm{\Gamma}(\{\underline{p}\},\mu)$ to distinguish the explicit scale dependence from the implicit one induced via the running coupling, we obtain
\begin{equation}\label{reslnZ}
   \ln\bm{Z}(\epsilon,\{\underline{p}\},\mu)
   = \int\limits_0^{\alpha_s} \frac{d\alpha}{\alpha}\,
    \frac{1}{2\epsilon-\beta(\alpha)/\alpha} 
    \Bigg[ \bm{\Gamma}(\{\underline{p}\},\mu,\alpha) 
    + \int\limits_0^\alpha \frac{d\alpha'}{\alpha'}\,
    \frac{\Gamma'(\alpha')}{2\epsilon-\beta(\alpha')/\alpha'}
    \Bigg] \,,
\end{equation}
where $\alpha_s\equiv\alpha_s(\mu)$, and we have defined
\begin{equation}\label{Gampr}
   \Gamma'(\alpha_s) 
   \equiv \frac{\partial}{\partial\ln\mu}\,
   \bm{\Gamma}(\{\underline{p}\},\mu,\alpha_s) 
   = - \gamma_{\rm cusp}(\alpha_s)\,\sum_i\,C_i \,.
\end{equation}
Note that this is a momentum-independent function, which is diagonal in color space. We have used that, when acting on color-singlet states, the unweighted sum over color generators can be simplified, because relation (\ref{singlet}) implies that
\begin{equation}\label{colorrel}
   \sum_{(i,j)}\,\bm{T}_i\cdot\bm{T}_j
   = - \sum_i\,\bm{T}_i^2 = - \sum_i\,C_i \,.
\end{equation}
This relation can be used in our case, because the scattering amplitudes are color conserving. Note that a different but equivalent form of relation (\ref{reslnZ}) has been given in \cite{Becher:2009cu}. 

It is understood that the result (\ref{reslnZ}) must be expanded in powers of $\alpha_s$ with $\epsilon$ treated as a fixed ${\cal O}(\alpha_s^0)$ quantity. Up to three-loop order this yields
\begin{eqnarray}\label{lnZ}
   \ln\bm{Z} 
   &=& \frac{\alpha_s}{4\pi} 
    \left( \frac{\Gamma_0'}{4\epsilon^2}
    + \frac{\bm{\Gamma}_0}{2\epsilon} \right) 
    + \left( \frac{\alpha_s}{4\pi} \right)^2 \! 
    \left[ - \frac{3\beta_0\Gamma_0'}{16\epsilon^3} 
    + \frac{\Gamma_1'-4\beta_0\bm{\Gamma}_0}{16\epsilon^2}
    + \frac{\bm{\Gamma}_1}{4\epsilon} \right] \\
   &&\!\!\!\mbox{}+ \left( \frac{\alpha_s}{4\pi} \right)^3\!
    \Bigg[ \frac{11\beta_0^2\,\Gamma_0'}{72\epsilon^4}
    - \frac{5\beta_0\Gamma_1' + 8\beta_1\Gamma_0' 
            - 12\beta_0^2\,\bm{\Gamma}_0}{72\epsilon^3} 
   + \frac{\Gamma_2' - 6\beta_0\bm{\Gamma}_1 
                - 6\beta_1\bm{\Gamma}_0}{36\epsilon^2}
    + \frac{\bm{\Gamma}_2}{6\epsilon} \Bigg] 
    + {\cal O}(\alpha_s^4) , \nonumber
\end{eqnarray}
where we have expanded the anomalous dimensions and $\beta$-function as
\begin{equation}\label{Gbexp}
   \bm{\Gamma} = \sum_{n=0}^\infty\,\bm{\Gamma}_n 
    \left( \frac{\alpha_s}{4\pi} \right)^{n+1} , \quad
   \Gamma' = \sum_{n=0}^\infty\,\Gamma'_n 
    \left( \frac{\alpha_s}{4\pi} \right)^{n+1} , \quad
   \beta = -2\alpha_s\,\sum_{n=0}^\infty\,\beta_n 
    \left( \frac{\alpha_s}{4\pi} \right)^{n+1} .
\end{equation}
Exponentiating the result (\ref{lnZ}) and taking into account that the different expansion coefficients $\bm{\Gamma}_n$ commute, it is straightforward to derive an explicit expression for $\bm{Z}$. For the convenience of the reader, we present the result along with the relevant expansion coefficients of the anomalous dimensions in  Appendix~A. Note that the highest pole in the ${\cal O}(\alpha_s^n)$ term of $\ln\bm{Z}$ is $1/\epsilon^{n+1}$, instead of $1/\epsilon^{2n}$ for the $\bm{Z}$-factor itself. The exponentiation of the higher pole terms was observed previously in \cite{Sterman:2002qn}. 

The IR singularities of two-loop scattering amplitudes were first predicted by Catani a decade ago \cite{Catani:1998bh}. The one- and two-loop coefficients of our $\bm{Z}$-matrix are closely related to his subtraction operators $\bm{I}^{(1)}$ and $\bm{I}^{(2)}$. Catani's  formula states that the product 
\begin{equation}
   \left[ 1 - \frac{\alpha_s}{2\pi}\,\bm{I}^{(1)}(\epsilon)
   - \left( \frac{\alpha_s}{2\pi} \right)^2 \bm{I}^{(2)}(\epsilon)
   + \dots \right] |{\cal M}_n(\epsilon,\{\underline{p}\})\rangle
\end{equation}
is free of IR poles through ${\cal O}(\alpha_s^2)$. The subtraction operators $\bm{I}^{(n)}(\epsilon)\equiv\bm{I}^{(n)}(\epsilon,\{\underline{p}\},\mu)$ are defined as
\begin{eqnarray}\label{I2}
\begin{aligned}
   \bm{I}^{(1)}(\epsilon) 
   &= \frac{e^{\epsilon\gamma_E}}{\Gamma(1-\epsilon)}\,
    \sum_i \left( \frac{1}{\epsilon^2} 
    - \frac{\gamma_0^i}{2\epsilon}\,\frac{1}{\bm{T}_i^2} \right)
    \sum_{j\neq i}\,\frac{\bm{T}_i\cdot\bm{T}_j}{2}
    \left( \frac{\mu^2}{-s_{ij}} \right)^\epsilon , \\
   \bm{I}^{(2)}(\epsilon)  
   &= \frac{e^{-\epsilon\gamma_E}\,\Gamma(1-2\epsilon)}%
          {\Gamma(1-\epsilon)} 
    \left( \frac{\gamma_1^{\rm cusp}}{8}
     + \frac{\beta_0}{2\epsilon} \right) 
    \bm{I}^{(1)}(2\epsilon) 
    - \frac12\,\bm{I}^{(1)}(\epsilon)  
    \left( \bm{I}^{(1)}(\epsilon) + \frac{\beta_0}{\epsilon} \right)  
    + \bm{H}_{\rm R.S.}^{(2)}(\epsilon) \,.
\end{aligned}
\end{eqnarray}
The conditions linking these objects to ours are
\begin{equation}\label{consistency}
   2\bm{I}^{(1)} \stackrel{!}{=} \bm{Z}_1 + \mbox{finite} \,,
    \qquad
   4\bm{I}^{(2)} \stackrel{!}{=} \bm{Z}_2 - 2\bm{I}^{(1)} \bm{Z}_1
    + \mbox{finite} \,,
\end{equation}
where $\bm{Z}_n$ denotes the coefficient of $(\alpha_s/4\pi)^n$ in the $\bm{Z}$-factor. The first relation is indeed satisfied. The second one can be used to derive an explicit expression for the quantity $\bm{H}_{\rm R.S.}^{(2)}$ encoding the genuine two-loop coefficient of the $1/\epsilon$ pole in (\ref{I2}), which was not obtained in \cite{Catani:1998bh}. We find\footnote{
The term in the third line was missing in the original version of our paper. This extra contribution was discussed in Appendix D of \cite{MertAybat:2006mz} and the form shown here was first given in \cite{Becher:2013vva}.}
\begin{equation}\label{H2}
\begin{split}
   \bm{H}_{\rm R.S.}^{(2)}(\epsilon) 
   &= \frac{1}{16\epsilon}\,\sum_i \bigg( \gamma_1^i 
    - \frac14\,\gamma_1^{\rm cusp}\,\gamma_0^i
    + \frac{\pi^2}{16}\,\beta_0\,\gamma_0^{\rm cusp}\,C_i \bigg) \\
   &\quad\mbox{}+ \frac{if^{abc}}{24\epsilon}\,
   \sum_{(i,j,k)} \bm{T}_i^a\,\bm{T}_j^b\,\bm{T}_k^c\,
    \ln\frac{-s_{ij}}{-s_{jk}} \ln\frac{-s_{jk}}{-s_{ki}} 
    \ln\frac{-s_{ki}}{-s_{ij}}  \\
     &   
     \quad\mbox{} - \frac{if^{abc}}{128\epsilon} \,\gamma_0^\text{cusp}
\sum_{(i,\,j,\,k)}\bT_i^a\,\bT_j^b\,\bT_k^c\;
\bigg( \frac{\gamma_0^i}{C_i} - \frac{\gamma_0^j}{C_j} \bigg)
\ln\frac{-s_{ij}}{-s_{jk}}\,\ln\frac{-s_{ki}}{-s_{ij}}\,,
\end{split}
\end{equation}
which apart from the last two
 terms is diagonal in color space and universal in the sense that it is a sum over contributions from each individual parton. Note that only the first term in this result is of a form suggested by (\ref{reslnZ}). The remaining terms in the first line  arise because the two-loop corrections involving the cusp anomalous dimension or the $\beta$-function are not implemented in an optimal way in (\ref{I2}). More importantly, the terms in the second and third lines of (\ref{H2}) arise only because the operator $\bm{I}^{(1)}$ in \cite{Catani:1998bh} is not defined in a minimal subtraction scheme, but also includes ${\cal O}(\epsilon^n)$ terms with $n\ge 0$.  As a result, the antisymmetric terms in the product $\bm{I}^{(1)}\bm{Z}_1$ in the second relation in (\ref{consistency}) contain the structure
\begin{equation}\label{H2extraalt}
   \frac{1}{16\epsilon}\,\sum_{(i,j)}\,\sum_{(k,l)}\,
   \ln\frac{\mu^2}{-s_{ij}}\,\ln^2\frac{\mu^2}{-s_{kl}}\,
   \big[ \bm{T}_i\cdot\bm{T}_j, \bm{T}_k\cdot\bm{T}_l \big] \,,
\end{equation}
as well as a second term with the same color commutator multiplied by a single logarithm $\ln\frac{\mu^2}{-s_{kl}}$ times $\gamma_0^k/C_k$. After some algebraic simplifications these reduce to the expressions shown in the second and third lines of (\ref{H2}). 
 Our result for $\bm{H}_{\rm R.S.}^{(2)}$ agrees with the findings of \cite{MertAybat:2006mz} and confirms a conjecture for the form of $\bm{H}_{\rm R.S.}^{(2)}$ for a general $n$-parton amplitude made in \cite{Bern:2004cz}. Note that the last two terms in (\ref{H2}) are only non-zero for four or more partons.
Due to color conservation the three-parton case involves only two independent generators $\bm{T}_1$ and $\bm{T}_2$, which is not sufficient to obtain a completely antisymmetric structure to contract with $f^{abc}$ in (\ref{H2}) or, equivalently, to get an non-zero commutator term in (\ref{H2extraalt}). Using momentum conservation,  one can furthermore show that the last term in (\ref{H2})  can only be present for amplitudes with more than four external legs \cite{MertAybat:2006mz}. 

In Section~\ref{sec:2loop} we will show that the non-trivial color and momentum structure in the second line (\ref{H2}) is incompatible with constraints derived from soft-collinear factorization, and thus it cannot appear at any order in the anomalous-dimension matrix, from which the $\bm{Z}$-factor is derived.

Our expressions (\ref{lnZ}) and (\ref{H2}) reproduce all known results for the two-loop $1/\epsilon^n$ poles of on-shell scattering amplitudes in massless QCD. In addition to the on-shell quark and gluon form factors, these include $e^+e^-\to\bar qqg$ \cite{Garland:2001tf} as well as all four-point amplitudes of quarks and gluons \cite{Anastasiou:2000kg,Anastasiou:2000ue,Anastasiou:2001sv,Glover:2001af,Bern:2002tk,Anastasiou:2002zn, Bern:2003ck}. At the three-loop level, only the IR divergences of the quark and gluon form factors are known for the QCD case \cite{Moch:2005id,Moch:2005tm,Baikov:2009bg}. For ${\cal N}=4$ SYM in the planar limit, on the other hand, the four-point functions are known up to four-loop order \cite{Anastasiou:2003kj,Anastasiou:2003kj}, and they also agree with our result.

An interesting alternative approach to the problem of IR singularities of on-shell amplitudes was developed in \cite{Sterman:2002qn}, where the authors exploited the factorization properties of scattering amplitudes \cite{Sen:1982bt,Kidonakis:1998bk,Kidonakis:1998nf} along with IR evolution equations familiar from the analysis of the Sudakov form factor \cite{Magnea:1990zb}. They recovered Catani's result (\ref{I2}) at two-loop order and related the coefficient of the unspecified $1/\epsilon$ pole term to a soft anomalous-dimension matrix, which was unknown at the time. They also explained how their method could be extended beyond two-loop order. The two-loop soft anomalous-dimension matrix was later calculated in \cite{MertAybat:2006wq,MertAybat:2006mz}. In very recent work, Gardi and Magnea have pushed this approach further and derived a set of constraint relations for the soft-anomalous dimension matrix, which hold to all orders in perturbation theory \cite{Gardi:2009qi}. We will comment later on the relations between our analysis and their work.

\section{Anomalous dimensions of $\bm{n}$-jet SCET operators}
\label{sec:SCET}

\subsection{Basic elements of SCET}
\label{sec:SCETbasics}

To analyze a hard-scattering process involving energetic particles propagating along the directions of unit three-vectors $\hat{\bm{n}}_i$ in SCET, we introduce two light-like reference vectors $n_i=(1,\hat{\bm{n}}_i)$ and $\bar n_i= (1,-\hat{\bm{n}}_i)$ for each direction, so that $n_i\cdot\bar n_i =2$. The effective theory then contains a set of collinear quark and gluon fields for each direction of large momentum flow. These describe partons with large energies $E_i\sim \sqrt{\hat s}$ associated with a jet of small invariant mass $M$. The small ratio of these scales, $\lambda=M^2/\hat{s}$, serves as the expansion parameter of the effective theory. The components of the momenta $p_c$ of the collinear quark and gluon fields $\chi_i$  and $ {\cal A}_{ i}^\mu$ associated with the $i$-th jet direction scale as
\begin{equation}\label{collscaling}
   \mbox{$i$-collinear:} \quad
   (\bar n_i\cdot p_c, n_i\cdot p_c, p_{c\perp}^\mu) 
   \sim (1, \lambda, \sqrt{\lambda}) \,,
\end{equation}
such that $p_c^2\sim\lambda\sim M^2$. The $\perp$ components are defined to be perpendicular to both $n_i$ and $\bar n_i$. Via the equation of motion, the scaling of the momenta also implies a scaling for the spin components of the fields. In the case of collinear fermions, it implies that two of the four components of the Dirac spinor field are power suppressed. These can be integrated out, after which the field fulfills the condition $\nslash_i\chi_i=0$. For the collinear gluon field, it implies that leading-power operators only depend on ${\cal A}_{i\perp}^\mu$.

In the absence of soft interactions, each collinear sector of the theory is equivalent to the original QCD Lagrangian \cite{Beneke:2002ph}. This is not surprising, since we can imagine going into the rest frame of any given jet, and the interactions among the partons of the jet will then be the same as in ordinary QCD. The particular scaling of the fields (\ref{collscaling}) is dictated by external kinematics, or more concretely by the source terms which generate them. The purely collinear SCET Lagrangian is thus simply given by $n$ copies of the ordinary QCD Lagrangian, and the effective-theory fields $\chi_i$ and ${\cal A}_{i\perp}^\mu$ are related to the usual quark and gluon fields via
\begin{equation}
   \chi_i(x) = W_i^\dagger(x)\,\frac{\nslash_i\nbslash_i}{4}\,
    \psi_i(x) \,, \qquad 
   {\cal A}_\perp^\mu(x) = W_i^\dagger(x)\,[iD_\perp^\mu\,W_i(x)] \,.
\end{equation}
The $i$-collinear Wilson lines
\begin{equation}\label{eq:Whc}
   W_{i}(x) = {\rm\bf P}\,\exp\left(ig\int_{-\infty}^0\!ds\,
   \nb_i\cdot A_i(x+s\nb_i) \right)
\end{equation}
ensure that these fields are invariant under collinear gauge transformations in each sector \cite{Bauer:2000yr,Bauer:2001yt}. The symbol $\bf{P}$ indicates path ordering, and the conjugate Wilson line $W_i^\dagger$ is defined with the opposite ordering prescription. 

In addition to collinear fields for each direction, the effective theory contains a single set of soft fields, which interact with all types of collinear fields. All components of the momenta $p_s$ carried by these fields scale as
\begin{equation}
   \mbox{soft:} \quad p_s^\mu\sim \lambda \,.
\end{equation}
This scaling is such that one can associate a soft parton with any of the $n$ jets without parametrically changing the invariant mass of the jet, because $(p_c+p_s)^2\sim\lambda$. The soft fields can thus mediate low-energy interactions between different collinear fields. However, at leading power this interaction is very simple: it can be obtained from the substitution $A_i^\mu(x)\to A_i^\mu(x)+\frac{\nb_i^\mu}{2}\,n_i\cdot A_s(x_-)$, or
\begin{equation}\label{eq:subst}
   {\cal A}_{i}^\mu(x)\to {\cal A}_{i}^\mu(x) + \frac{\nb_i^\mu}{2}\,
   W^\dagger_{i}(x)\,g\,n_i\cdot A_s(x_-)\,W_{i}(x) \,,
\end{equation}
in each of the collinear Lagrangians, where $x_-=(\bar n_i\cdot x)\,n_i/2 $. Only the $n_i\cdot A_s$ component of the soft gluon field enters in this relation, since all other components are power suppressed compared to the collinear fields. The peculiar $x$-dependence of the gluon field is a consequence of the multipole expansion \cite{Beneke:2002ph,Beneke:2002ni}, which implies that in interactions of collinear and soft fields one should perform a derivative expansion of the form
\begin{equation}
   \left[ \phi_c(x) \right]^2 \phi_s(x) 
   = \left[ \phi_c(x)\right]^2 \left[ \phi_s(x_-) 
   + x_\perp\cdot\partial_\perp\,\phi_s(x)|_{x=x_-} + \dots \right] .
\end{equation}
The first-derivative term is suppressed by $\sqrt\lambda$, because $x_\perp\sim 1/\sqrt\lambda$, while $\partial_\perp$ acting on a soft field counts as ${\cal O}(\lambda)$. 

The substitution (\ref{eq:subst}) gives rise to an eikonal interaction of soft gluons with collinear fermion fields,
\begin{equation}
   {\cal L}_{c_i+s} = \bar\chi_i(x)\,\frac{\nbslash_i}{2}\,
   n_i\cdot A_s(x_-)\,\chi_i(x) \,.
\end{equation}
This interaction can be represented in terms of soft Wilson lines. Redefining the quark and gluon fields as
\begin{equation}\label{decouple}
\begin{aligned}
   \chi_i(x) &= S_i(x_-)\,\chi_i^{(0)}(x) \,, \\
   \bar\chi_i(x) &= \bar\chi_i^{(0)}(x)\,S_i^\dagger(x_-) \,, \\
   {\cal A}_{i\perp}^\mu(x) 
   &= S_i(x_-)\,{\cal A}_{i\perp}^\mu(x)\,S_i^\dagger(x_-) \,,
\end{aligned}
\end{equation}
where 
\begin{equation}\label{softwilson}
   S_i(x) = {\bf P} \exp\left( ig \int_{-\infty}^0\,dt\,
   n_i\cdot A_s^a(x+tn_i)\,t^a \right) ,
\end{equation}
eliminates the interaction ${\cal L}_{c_i+s}$ (including the pure-gluon terms). After this decoupling transformation \cite{Bauer:2001yt}, soft interactions manifest themselves as Wilson lines in operators built from collinear fields. The soft gluons do not couple to the spin of the collinear particles, and for the discussion that follows the spin degrees of freedom will be irrelevant. 

As written above the soft Wilson lines $S_i$ and $S_i^\dagger$ are color matrices defined in the fundamental representation of the gauge group. The transformations (\ref{decouple}) take on a universal form if we define a soft Wilson line $\bm{S}_i$ in analogy with (\ref{softwilson}), but with $t^a$ replaced by the color generator $\bm{T}_i^a$ in the appropriate representation for the $i$-th parton. Representing a generic collinear field as $(\phi_i)_{a_i}^{\alpha_i}(x)$ with color index $a_i$ and Dirac/Lorentz index $\alpha_i$, the soft interactions can then be decoupled from this field by the redefinition
\begin{equation}
   (\phi_i)_{a_i}^{\alpha_i}(x)
   = [ {\bm S}_i(x_-) ]_{a_i b_i}\, 
   {(\phi_i)^{(0)}}_{b_i}^{\alpha_i}(x) \,.
\end{equation}
Note that even anti-quarks transform according to this rule: in this case $\bm{T}_i^a=-(t^a)^T$, which translates into the anti-path ordering in (\ref{decouple}).

Hard interactions among the different jets are integrated out in the effective theory and absorbed into the Wilson coefficients of operators composed out of products of collinear and soft fields. Since additional soft fields in the SCET operators would lead to power suppression, the leading $n$-jet operators are built from $n$ collinear fields, one for each direction of large energy flow \cite{Bauer:2006qp,Bauer:2006mk}. The most general such operator with given particle content appears in the effective Hamiltonian
\begin{equation}
   {\cal H}_n^{\rm eff} = \int\!dt_1\,\dots\,dt_n\, 
   {\tilde{\cal C}}_{\alpha_1\dots\alpha_n}^{a_1\dots a_n}
    (t_1,\dots,t_n,\mu)\, 
   (\phi_1)_{a_1}^{\alpha_1}(x+t_1\bar n_1)\dots  
   (\phi_n)_{a_n}^{\alpha_n}(x+t_n\bar n_n) \,.
\end{equation}
Our notation is somewhat unusual, because the Wilson coefficients of these operators carry spin and color indices. Usually both operators and Wilson coefficients are chosen to be color-neutral Lorentz scalars. However, writing the operator in this form makes the connection to the color-space notation we use for the scattering amplitudes most transparent. Adopting this notation, the effective Hamiltonian for an $n$-jet process reads
\begin{equation}\label{hard}
   {\cal H}_n^{\rm eff} = \int\!dt_1\,\dots\,dt_n\,
   \langle O_n(\{\underline{t}\},\mu)|
   \tilde{\cal C}_n(\{\underline{t}\},\mu)\rangle \,,
\end{equation}
where $\mu$ is the scale at which the SCET operator is renormalized. An $n$-parton scattering amplitude is obtained by taking an on-shell partonic matrix element of this operator. In this step effective-theory loop integrals vanish in dimensional regularization, because they are scaleless. The on-shell matrix elements are therefore given by their tree-level values, consisting of products of on-shell spinors and polarization vectors defined through the relations
\begin{equation}
\begin{aligned}
   \langle 0|(\chi_j)_{\alpha}^{a}(t_j\bar n_j)|p_i;a_i,s_i\rangle 
   &= \delta_{ij}\,\delta_{a_i a}\,e^{-it_i\bar n_i\cdot p_i}\,
    u_{\alpha}(p_i,s_i) \,, \\
   \langle 0|({{\cal A}_j}_\perp)^a_\mu(t_j\bar n_j)|p_i;a_i,s_i
   \rangle
   &= \delta_{ij}\,\delta_{a_i a}\,e^{-it_i\bar n_i\cdot p_i}\,
    \epsilon_\mu(p_i,s_i) \,.
\end{aligned}
\end{equation}
Loop corrections to the scattering amplitude are encoded in the Wilson coefficients $\tilde{\cal C}_n(\{\underline{t}\},\mu)$. The integrations over $t_i$ in (\ref{hard}) produce the Fourier transforms ${\cal C}_n(\{\underline{p}\},\mu)$ of these coefficients, which after contraction with the spinors and polarization vectors arising when taking the tree-level matrix elements are in one-to-one correspondence with the scattering amplitudes \cite{Becher:2009cu}, as shown in (\ref{MnCn}).   
 
\subsection{Soft-collinear factorization and decoupling}
\label{sec:decomp}

To obtain the general form of the anomalous-dimension matrix $\bm{\Gamma}$ defined in (\ref{RGE}), we now derive a factorization theorem for the matrix elements of SCET operators. The first factorization step has already been achieved in (\ref{hard}), which separates hard from soft and collinear fluctuations. In a second step, we separate the collinear and soft degrees of freedom using the decoupling transformation (\ref{decouple}), which eliminates the leading-power interactions among soft and collinear fields. Since collinear fields from different sectors do not interact directly, this completely factorizes a matrix element into a soft part $\cal S$, given by a product of Wilson lines along the directions $n_i$, and a product of collinear matrix elements ${\cal J}_i$ for each direction.
 
RG invariance implies that the right-hand side of (\ref{hard}) must be independent of the renormalization scale. Denoting by $\bm{\Gamma}_h\equiv\bm{\Gamma}$ the anomalous-dimension matrix of the hard contributions contained in the Wilson coefficient functions ${\cal C}_n$ and by $\bm{\Gamma}_{c+s}$ the anomalous dimension associated with the collinear and soft contributions contained in the matrix elements of the SCET operators,\footnote{Following common practice, we define the anomalous dimensions of operators with the opposite sign compared to those for Wilson coefficients.} 
it follows that $\bm{\Gamma}_h=\bm{\Gamma}_{c+s}$. The decoupling transformation, which removes the interactions of collinear fields with soft gluons and absorbs them into Wilson lines \cite{Bauer:2001yt}, allows us to further decompose $\bm{\Gamma}_{c+s}=\bm{\Gamma}_c+\bm{\Gamma}_s$. There are no mixed soft-collinear contributions. The collinear piece $\bm{\Gamma}_c=\sum_i\Gamma_c^i$ is a sum over color-singlet single-particle contributions, because the fields belonging to different collinear sectors of SCET do not interact with one another. Hence, contributions to the anomalous dimension involving correlations between different partons only reside in the soft and hard contributions, $\bm{\Gamma}_s$ and $\bm{\Gamma}_h$, and they coincide.

\begin{figure}
\begin{center}
\psfrag{x}[b]{$n_1$}
\psfrag{y}[l]{$n_2$}
\psfrag{z}[]{$n_3$}
\psfrag{u}[]{$n_4$}
\psfrag{w}[]{$\phantom{ii}n_5$}
\psfrag{v}[b]{$n_n$}
\includegraphics[height=4.3cm]{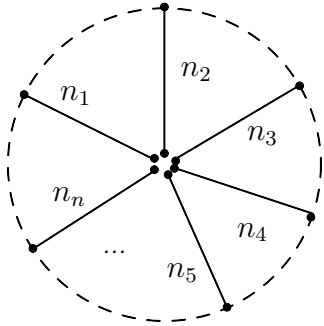}
\end{center}
\vspace{-4mm}
\caption{\label{fig:star}
Graphical representation of the soft operator ${\cal S}(\{\underline{n}\},\mu)$ corresponding to an $n$-parton scattering amplitude. The $n$ light-like Wilson lines start at the origin and run to infinity. The dots represent open color indices.}
\end{figure}

After the decoupling transformation the soft matrix element is a vacuum expectation value of $n$ light-like Wilson lines, one for each external parton in the associated color representation:
\begin{equation}\label{Slines}
   {\cal S}(\{\underline{n}\},\mu) 
   = \langle0|\bm{S}_1(0)\ldots\bm{S}_n(0) |0\rangle \,.
\end{equation}
As illustrated in Figure~\ref{fig:star}, this object is an operator in color space, with each $\bm{S}_i$ factor operating on the color indices of the $i$-th parton. Its renormalization properties are strongly constrained by the simplicity of soft gluon interactions, which only probe the direction of the Wilson lines and their color charge. When the color indices are contracted in color-singlet combinations, then ${\cal S}(\{n\},\mu)$ turns into products of closed Wilson loops, which touch or intersect each other at the origin. The renormalization properties of such Wilson loops have been studied extensively in the literature, see e.g.\ \cite{Dotsenko:1979wb,Polyakov:1980ca,Brandt:1981kf,Gatheral:1983cz,Frenkel:1984pz,Dorn:1986dt,Korchemsky:1987wg} and references therein. We will use several results obtained in these studies and generalize them to the case of the Wilson-line operator in (\ref{Slines}). We will also indicate where known properties of Wilson loops correspond to certain features of the effective theory and vice versa.

For on-shell amplitudes, the loop integrals in the effective theory have both IR and UV divergences and vanish in dimensional regularization. This makes the correspondence between the Wilson coefficients in (\ref{hard}) and the amplitudes manifest. However, because of the cancellations between UV and IR poles, we cannot use on-shell amplitudes to obtain the anomalous dimensions of the SCET operators. To separate out the UV divergences we need to consider IR-finite quantities. The simplest possibility is to consider slightly off-shell $n$-parton amputated Green's functions $G_n(\{\underline{p}\})$. However, in this case we encounter a subtlety. While the off-shell Green's function in QCD and SCET are IR finite, this is no longer the case after the field redefinition (\ref{decouple}). Field redefinitions leave ``physical'' quantities such as on-shell matrix elements unchanged, but they can change the off-shell behavior of fields. To calculate the anomalous dimensions perturbatively from off-shell Green's functions, one should  use the original, non-decoupled fields.\footnote{Alternatively, one can perform the calculations using a different IR regulator, e.g.\ by considering finite-length Wilson lines with $n_i^2\neq 0$  \cite{Korchemskaya:1992je}.} For the case of the quark form factor, the corresponding one-loop calculation in the effective theory was performed in \cite{Becher:2003kh}. Generalizing this result to $n$-point functions, we find for the UV poles of the jet and soft functions (normalized to 1 at tree level)
\begin{equation}\label{JS1loop}
\begin{aligned}
   {\cal J}_q(p^2,\mu) 
   &= 1 + \frac{\alpha_s}{4\pi}\,C_F
   \left( \frac{2}{\epsilon^2} + \frac{2}{\epsilon}\,
    \ln\frac{\mu^2}{-p^2} + \frac{3}{2\epsilon} \right)
    + {\cal O}(\epsilon^0) \,, \\
   {\cal J}_g(p^2,\mu) 
   &= 1 + \frac{\alpha_s}{4\pi} \left[
    C_A \left( \frac{2}{\epsilon^2} + \frac{2}{\epsilon}\,
    \ln\frac{\mu^2}{-p^2} \right) + \frac{\beta_0}{2\epsilon} \right]
    + {\cal O}(\epsilon^0) \,, \\
   {\cal S}(\{\underline{p}\},\mu) 
   &= 1 + \frac{\alpha_s}{4\pi}\,\sum_{(i,j)}\,
    \frac{\bm{T}_i\cdot\bm{T}_j}{2} 
    \left( \frac{2}{\epsilon^2} + \frac{2}{\epsilon}\, 
    \ln\frac{-\sigma_{ij}\,n_i\cdot n_j\,\nb_i\cdot p_i\,
             \nb_j\cdot p_j\,\mu^2}{2(-p_i^2)(-p_j^2)} \right)
    + {\cal O}(\epsilon^0) \,.
\end{aligned}
\end{equation}
The jet functions are color-diagonal. The reference vectors $n_i$ are defined such that up to power corrections $p_i=E_i\,n_i$, which implies that at leading power $\frac12\,\sigma_{ij}\,n_i\cdot n_j\,\nb_i\cdot p_i\,\nb_j\cdot p_j=2\sigma_{ij}\,p_i\cdot p_j=s_{ij}$. The one-loop divergences of the complete effective-theory $n$-particle matrix elements are thus given by
\begin{equation}
   {\cal S}(\{\underline{p}\},\mu)\,\prod_i\,{\cal J}_i(p_i^2,\mu) 
   = 1 - \frac{\alpha_s}{4\pi}\,\bigg[ \sum_{(i,j)}\,
   \frac{\bm{T}_i\cdot\bm{T}_j}{2} \left( \frac{2}{\epsilon^2} 
   + \frac{2}{\epsilon}\,\ln\frac{\mu^2}{-s_{ij}} \right)
   + \sum_i\,\frac{\gamma_0^i}{2\epsilon} + {\cal O}(\epsilon^0)
   \bigg] \,.
\end{equation}
The UV divergences of the SCET operator matrix elements are equal and opposite to those of the corresponding Wilson coefficient. They are thus given by minus the one-loop coefficient of $\ln\bm{Z}$ in (\ref{lnZ}). The one-loop coefficients of the jet-function anomalous dimensions are obtained from (\ref{JS1loop}) as $\gamma_0^q=-3C_F$ and $\gamma_0^g=-\beta_0$. Interestingly, the dependence on the off-shellness cancels in the divergent part of the full matrix elements, which only depend on the hard scales $s_{ij}$. This cancellation has to occur, because the anomalous dimensions of the operators cannot depend on low-energy scales such as $p_i^2$. Otherwise the scale dependence of the operators could not possibly cancel against the scale dependence of the hard Wilson coefficients, which are insensitive to IR scales. As we will explain below, the factorization property $\bm{\Gamma}_h=\bm{\Gamma}_s+\sum_i\Gamma_c^i$ places a strong constraint on the form of the anomalous dimension.

\subsection{Color-symmetrized soft gluon attachments}
\label{sec:softattachments}

\begin{figure}
\psfrag{a}[b]{$n_1$}
\psfrag{b}[b]{$n_2$}
\psfrag{c}[b]{$n_l$}
\psfrag{x}[b]{$~\mu,a$}
\psfrag{u}[b]{$~\nu,b$}
\psfrag{y}[b]{$k_1$}
\psfrag{v}[b]{$k_2$}
\psfrag{z}[]{$\hspace{2.9cm} 
 = \displaystyle\sum_i\,g\,\frac{n_i^\mu}{n_i\cdot k_1}\,\bm{T}_i^a$}
\psfrag{g}[]{$\hspace{6.5cm}
 = \displaystyle\sum_{i,j}\,\frac{g^2}{2!}\,
 \frac{n_i^\mu\,n_j^\nu}{n_i\cdot k_1\,n_j\cdot k_2}\,
 \{\bm{T}_i^a,\bm{T}_j^b\} + \dots$}
\hspace{2cm}\includegraphics[width=5.5cm]{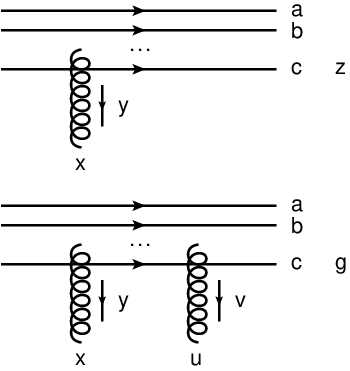}
\vspace{-3mm}
\caption{\label{fig:rules}
Feynman rules for the sums over all possible attachments of one (top) or two (bottom) soft gluons to a set of Wilson lines. The indices $i,j$ run from 1 to $l$.}
\end{figure}

Of special importance to our analysis in Section~\ref{sec:arguments} will be the fact that the interactions of soft gluons with collinear partons take on a particularly simple form after a symmetrization of color matrices has been performed. To derive the form of this relation, consider the sums over all possible attachments of one or two soft gluons to a set of $l$ Wilson lines $\bm{S}_i(0)$ with tangent vectors $n_i$, such as those contained in the soft operator in (\ref{Slines}). Using the Feynman rules of SCET, we find the simple expressions shown in Figure~\ref{fig:rules}, where the dots in the second relation represent an anti-symmetric color structure containing the commutator $[\bm{T}_i^a,\bm{T}_i^b]$ for a single parton index. Such a structure can be reduced to a single color generator using the Lie algebra of the gauge group. The important point to note about this result is that for the {\em symmetric\/} color structure the momentum dependence is the same irrespective of whether the soft gluons attach to the same or to different Wilson lines. In the two-gluon case, this happens because the two possible attachments to a single Wilson line yield
\begin{equation}
   \frac{\bm{T}_i^b\,\bm{T}_i^a}{n_i\cdot(k_1+k_2)\,n_i\cdot k_2}
   + \frac{\bm{T}_i^a\,\bm{T}_i^b}{n_i\cdot(k_1+k_2)\,n_i\cdot k_1}
   \,,
\end{equation}
which after symmetrization gives rise to the structure shown in Figure~\ref{fig:rules}.

The Feynman rules given in the figure generalize in an obvious way to the case of more than two soft gluons. The reason is that the symmetrization in color generators eliminates the need for the path-ordering symbol in the definition (\ref{softwilson}) of the soft Wilson lines. Introducing the notation
\begin{equation}\label{symprod}
   \left( \bm{T}_{i_1}^{a_1}\dots\,\bm{T}_{i_n}^{a_n} \right)_+
   \equiv \frac{1}{n!}\,\sum_{\sigma\in S_n}\,
   \bm{T}_{i_{\sigma(1)}}^{a_{\sigma(1)}}\dots\,
   \bm{T}_{i_{\sigma(n)}}^{a_{\sigma(n)}}
\end{equation}
for the symmetric product of $n$ color matrices, where $S_n$ is the set of permutations of $n$ objects, we obtain for the general case of $m$ soft gluons attached to a set of Wilson lines the Feynman rule
\begin{equation}
   \sum_{i_1,\dots,i_m} g^m\,
   \frac{n_{i_1}^{\mu_1}\dots\,n_{i_m}^{\mu_m}}%
        {n_{i_1}\cdot k_1\dots\,n_{i_m}\cdot k_m}\,
   \left( \bm{T}_{i_1}^{a_1}\dots\,\bm{T}_{i_m}^{a_m} \right)_+ 
   + \dots \,.
\end{equation}
Since color generators acting of different partons commute, the symmetric product in (\ref{symprod}) symmetrizes the color matrices acting on each individual parton. For instance, we have for different indices $i$, $j$, $k$
\begin{equation}
   \left( \bm{T}_i^a\,\bm{T}_j^b\,\bm{T}_i^c\,\bm{T}_k^d\,
    \bm{T}_k^e \right)_+
   = \big( \bm{T}_i^a\,\bm{T}_i^c \big)_+ \bm{T}_j^b 
    \left( \bm{T}_k^d\,\bm{T}_k^e \right)_+ \,.
\end{equation}

\section{Constraints from soft-collinear factorization}
\label{sec:arguments}

We will now summarize the arguments that have led us to propose the conjecture (\ref{magic}). Approaching the problem from the perspective of effective field theory adds one crucial element to the discussion. This is the realization that the IR-divergent terms in the scattering amplitudes must obey some restrictive constraints, which do not apply to the finite remainders. The scattering amplitudes are in general complicated functions of the kinematical invariants $s_{ij}$ as well as of the color, spin, and polarization quantum numbers of the external partons. The different $s_{ij}$ variables are assumed to be hard scales of the same order of magnitude, so that the ratio of any two such variables is an ${\cal O}(1)$ quantity. In principle, arbitrary functions of combinations of such ratios can arise in the expressions for the scattering amplitudes. The situation is, however, very different for the IR-singular terms in the amplitudes. RG invariance of the effective theory requires that the anomalous dimensions of the hard matching coefficients $|{\cal C}_n(\{\underline{p}\},\mu)\rangle$, which according to (\ref{MnCn}) correspond to the on-shell scattering amplitudes, must be decomposable into sums of collinear and soft contributions. This requires a rewriting of the hard momentum variables $s_{ij}$ in terms of soft and collinear variables. The very fact that such a rewriting must exist restricts the functional dependence of the anomalous dimension on the $s_{ij}$ variables to be either single logarithmic or in the form of certain conformal cross ratios, which will be defined below. Moreover, the structure of the effective theory enforces that terms depending on the collinear variables cannot lead to correlations between different partons and must be diagonal in color space. Correlations can only arise through soft gluon exchange. The universal structure of these interactions implies that any dependence on the identity of the external partons can only arise via their momenta and color charges, but not through spin information. We will also discuss constraints on the color structure of the soft anomalous-dimension matrix implied by the non-abelian exponentiation theorem and other considerations.

\subsection{Renormalization of Wilson loops}

A well-known property of Wilson loops is that they require UV subtractions beyond the renormalization of the coupling constant in cases where the integration path is not smooth, but contains one or more singular points \cite{Dotsenko:1979wb,Polyakov:1980ca,Brandt:1981kf}. These divergences can be removed multiplicatively. The simplest case is that of a Wilson loop with a single cusp, i.e., a point where the tangent vector changes its direction abruptly. If the cusp is formed by two time-like segments with tangent vectors $n_1$ and $n_2$, then these UV divergences are removed by a factor $Z(\beta_{12})$, which is a function of the hyperbolic cusp angle $\beta_{12}$ defined by 
\begin{equation}\label{cuspangle}
   \cosh\beta_{12} = \frac{n_1\cdot n_2}{\sqrt{n_1^2\,n_2^2}} \,,
\end{equation}
where for simplicity we have assumed that $n_1$ points into the cusp and $n_2$ points out of it. If the Wilson loop has more than one cusp, then each of them contributes an analogous $Z$-factor.

\begin{figure}
\begin{center}
\includegraphics[height=2.0cm]{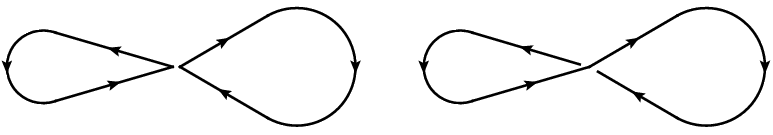}
\end{center}
\vspace{-4mm}
\caption{\label{fig:Wilsonloops}
Color-singlet contractions of four Wilson lines in the fundamental representation. The resulting Wilson-loop operators mix under renormalization.}
\end{figure}

A more complicated situation arises if, as in our case, different Wilson lines cross each other at a point. Then Wilson loops tracing out the same space-time curves except for the cross point mix under renormalization. An example are the two Wilson-loop operators shown in Figure~\ref{fig:Wilsonloops}, which illustrates this fact for the case of a four-jet operator corresponding to $q\bar q\to q\bar q$ scattering. The renormalization factor $\bm{Z}(\{\underline{\beta}\})$ is then a matrix on the space of such Wilson loops, which depends on the set of all hyperbolic angles formed by the tangent vectors at the cross point \cite{Brandt:1981kf}. Generalizing these results to our case, where the Wilson-line operators are matrices in color space as shown in (\ref{Slines}), the renormalization factor must be promoted to a soft matrix $\bm{Z}_s$ acting on the product space of the color representations of the $n$ partons.

RG invariance implies that the renormalization factor can be constructed in the usual way from a soft anomalous-dimension matrix $\bm{\Gamma}_s$. For the case of a single cusp, the two-loop expression for the anomalous dimension was first obtained in \cite{Korchemsky:1987wg,Korchemsky:1988hd}. 

\subsection{Non-abelian exponentiation theorem}

The structure of the soft anomalous-dimension matrix is restricted by the non-abelian exponentiation theorem \cite{Gatheral:1983cz,Frenkel:1984pz}, which implies that purely virtual amplitudes in the eikonal approximation (i.e., with only soft gluon interactions taken into account) can be written as exponentials of simpler quantities, which only receive contributions from Feynman graphs whose color weights are ``maximally non-abelian'' (or ``color-connected''). Applied to our case, it follows that the logarithm of the soft $\bm{Z}_s$-factor, and with it the soft anomalous-dimension matrix $\bm{\Gamma}_s$, only receives such contributions.

\begin{figure}
\begin{center}
\includegraphics[height=1.6cm]{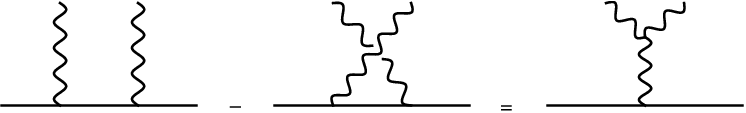}
\end{center}
\vspace{-4mm}
\caption{\label{fig:Jacobi}
Diagrammatic form of the Lie commutator relation. Gluons are drawn as wavy lines in order to distinguish color-weight graphs from Feynman diagrams.}
\end{figure}

\begin{figure}[t]
\begin{center}
\vspace{3mm}
\includegraphics[width=10.0cm]{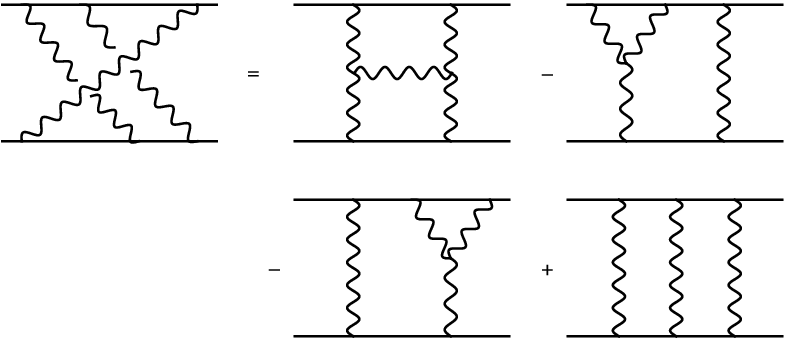}
\end{center}
\vspace{-4mm}
\caption{\label{fig:cweb}
Decomposition of a web into a sum of products of connected webs. The non-abelian exponentiation theorem states that only the single connected web shown in the first graph on the right contributes to the color weights in the exponent of the amplitude.}
\end{figure}

We follow the diagrammatic approach to the non-abelian exponentiation theorem developed in \cite{Frenkel:1984pz}, since it is more explicit and intuitive than the iterative construction presented in \cite{Gatheral:1983cz}. To each Feynman diagram we assign a color-weight diagram, in which vertices are replaced by color matrices $(t^a)_{ij}$ or structure constants $-if^{abc}$ (or, more generally, by generators $\bm{T}^a$ in the appropriate representation of the gauge group), and propagators by $\delta^{ij}$ for quarks and $\delta^{ab}$ for gluons or ghosts. Color diagrams may be related to one another by use of the Lie algebra relation $[\bm{T}^a,\bm{T}^b]=if^{abc}\,\bm{T}^c$, as illustrated in Figure~\ref{fig:Jacobi}. In the adjoint representation this is called the Jacobi identity. One defines a connected web as a connected set of gluon lines (possibly with internal fermion loops), not counting crossed lines as being connected. It has been shown in \cite{Frenkel:1984pz} that, using the Lie commutator relation, any color-weight diagram can be written as a sum over products of connected webs. The non-abelian exponentiation theorem implies that only {\em single connected webs\/} contribute to the color weights in the exponent. For the case of two Wilson lines this is illustrated with an example in Figure~\ref{fig:cweb}. In this special case, the above definitions imply that the single connected webs contain those diagrams that are two-particle (rainbow) irreducible diagrams with respect to the Wilson lines \cite{Korchemsky:1987wg}.

Note that in our case the gluons can be attached to more than two Wilson lines, provided there are more than two external partons. The notion of crossed lines is meaningful only for diagrams such as Figure~\ref{fig:cweb}, in which only two Wilson lines are connected. The reduction to connected webs needs to be generalized when more than two lines are connected by gluons in a diagram. In general, if one has a diagram with $n$ gluons connected to a given Wilson line $i$, one first symmetrizes the product of color matrices using the identity
\begin{equation}\label{reducecol}
   \bm{T}_{i}^{a_1}\dots\,\bm{T}_{i}^{a_n} 
   = \left( \bm{T}_{i}^{a_1}\dots\,\bm{T}_{i}^{a_n} \right)_+ 
   + \mbox{``commutator terms''.}
\end{equation}
The ``commutator terms'' involve one or more commutators of color generators and can thus be reduced to products of $(n-1)$ or fewer generators multiplied by structure constants. Repeatedly applying the identity (\ref{reducecol}), one can write any color structure as symmetric products of generators multiplied by structure constants. By performing the symmetrization on all legs, one splits each diagram into a number of webs, and the statement of non-abelian exponentiation is that only single connected webs contribute in the exponent. Since all attachments are fully symmetrized, the distinction between crossed and uncrossed lines is no longer relevant.

The fact that only single connected webs contribute to the logarithm of the $\bm{Z}_s$-factor (and hence to the anomalous dimension), while products of webs contribute to the $\bm{Z}_s$-factor itself, is in analogy to the usual structure of UV divergences in quantum field theory, as described by Zimmermann's forest formula. Formal arguments explaining the systematics of UV divergences for arbitrary Wilson loops can be found in  \cite{Brandt:1981kf}.

\subsection{Light-like Wilson lines}
\label{sec:singlelog}

For large values of the cusp angle $\beta_{12}$ in (\ref{cuspangle}), the anomalous dimension $\Gamma(\beta_{12})$ associated with a cusp (or cross point) grows linearly with $\beta_{12}$ \cite{Korchemsky:1987wg}, which in this case is approximately equal to $\ln(2n_1\cdot n_2/\sqrt{n_1^2\,n_2^2})$. In the limit where one or both segments forming the cusp approach a light-like direction, the cusp angle diverges ($\beta_{12}\to\infty$). In dimensional regularization this divergence gives rise to a single logarithm of the renormalization scale in the anomalous dimension. If both segments lie on the light-cone, then \cite{Korchemskaya:1992je}
\begin{equation}\label{singlelog}
   \Gamma(\beta_{12}) \stackrel{n_{1,2}^2\to 0}{\to} 
   \Gamma_{\rm cusp}^i(\alpha_s)\,\ln\frac{\mu^2}{\Lambda_s^2} 
   + \dots \,,
\end{equation}
where we refer to $\Gamma_{\rm cusp}^i(\alpha_s)$ as {\em the\/} cusp anomalous dimension in the color representation of parton $i$. Its two-loop expression was obtained long ago in \cite{Kodaira:1981nh,Catani:1988vd} and \cite{Korchemsky:1987wg,Korchemsky:1988hd}, while the three-loop result was derived in \cite{Moch:2004pa}. The above equation is formal and meant to show the dependence on the renormalization scale only. We will explain later how a soft scale $\Lambda_s$ with the proper dimensions appears in the argument of the logarithm.

In conventional applications of the RG, large (single) logarithms of scale ratios entering perturbative results for multi-scale problems can be resummed with the help of anomalous dimensions that are functions of the coupling constant, much like the $\beta$-function. This resums terms of the form $(\alpha_s L)^n$ in the perturbative series, where $L$ is the logarithm of the relevant scale ratio. However, the presence of overlapping soft and collinear singularities in on-shell scattering amplitudes of massless partons generates Sudakov double logarithms of the form $(\alpha_s L^2)^n$ in perturbation theory. They can be resummed with the help of anomalous dimensions which themselves contain a single logarithm $L$ of the large scale ratio. The logarithmic dependence of the anomalous dimension in (\ref{singlelog}) is an essential feature in this context. 

SCET is the appropriate effective field theory to formalize applications of the RG to observables sensitive to Sudakov double logarithms. It has been used in the past to derive the exact form of the anomalous dimensions of the two-jet operators for quarks and gluons, which were found to be of the form \cite{Manohar:2003vb,Becher:2006nr,Becher:2007ty,Ahrens:2008nc}
\begin{equation}\label{2jet}
   \Gamma_{\rm 2-jet} 
   = - \Gamma_{\rm cusp}^i(\alpha_s)\,\ln\frac{\mu^2}{-s} 
   + 2\gamma^i(\alpha_s) \,.
\end{equation}
Here $i=q$ for quarks and $i=g$ for gluons, and $s\equiv s_{12}$ is the invariant momentum transfer. From the perspective of the effective theory, the appearance of the cusp logarithm is, at first sight, perplexing. How can the anomalous dimension of an operator defined in an effective theory, in which the hard scale $s$ has been integrated out, remember the value of that scale? The answer to this puzzle was given in \cite{Becher:2003kh}, where it was explained that the hard scale is imprinted in the effective theory via the large rapidity that separates the rest frames of soft and collinear hadrons in a given physical process. This leads to a characteristic entanglement of the hard, collinear, and soft mass scales, the latter two of which are known to the effective theory. This correlation is such that $\mu_c^2\sim\mu_h\,\mu_s$, and the hard logarithm in relation (\ref{2jet}) can thus be rearranged in the symbolic form
\begin{equation}\label{decomp}
   \ln\frac{\mu^2}{\mu_h^2} 
   = 2\ln\frac{\mu^2}{\mu_c^2} - \ln\frac{\mu^2}{\mu_s^2} \,.
\end{equation}
Obviously, such a rearrangement of a hard contribution as a sum of collinear and soft contributions is only possible for functions  containing either constants or single logarithms of scale ratios, a point that was also emphasized in \cite{Manohar:2003vb}. We observe an interesting connection between RG invariance in SCET and a property of soft Wilson loops: RG invariance requires a single-logarithmic dependence of the anomalous dimension on the hard scale, because only then can this dependence be decomposed into dependences on collinear and soft scales. Relation (\ref{decomp}) then implies single-logarithmic dependence on the soft scale, in accordance with known renormalization properties of Wilson loops mentioned above, see (\ref{singlelog}).

That a decomposition of the form (\ref{decomp}) is indeed at work in the effective theory was demonstrated in \cite{Becher:2003kh} using the method of regions, by analyzing the collinear and soft contributions to the anomalous dimension (\ref{2jet}) separately. As we have explained in Section~\ref{sec:decomp}, in doing this it is necessary to either consider physical quantities, which are IR finite, or to
introduce an IR regulator scale in order to define the collinear and soft scales. Using a regulator introduces some arbitrariness and scheme dependence into the calculation of the individual contributions, which however vanishes in their sum (see also \cite{MertAybat:2006mz,Gardi:2009qi}). For concreteness, we introduce a small off-shellness $(-p_i^2)>0$ for the external partons, taking the limit $p_i^2\to 0$ wherever possible. The decomposition of a generic hard logarithm then reads
\begin{equation}\label{decomp2}
   \ln\frac{\mu^2}{-s_{ij}} 
   = \ln\frac{\mu^2}{-2\sigma_{ij}\,p_i\cdot p_j} 
   = \ln\frac{\mu^2}{-p_i^2} + \ln\frac{\mu^2}{-p_j^2} 
   - \ln\frac{-2\sigma_{ij}\,p_i\cdot p_j\,\mu^2}%
             {(-p_i^2)(-p_j^2)} \,.
\end{equation}
This is precisely the structure of collinear and soft logarithms found in \cite{Becher:2003kh}. Measuring all scales in units of the hard scale, we have the power counting $p_i\cdot p_j\sim 1$ for the hard scales, $p_i^2\sim p_j^2\sim\lambda$ for the collinear scales, and  $p_i^2\,p_j^2/p_i\cdot p_j\sim\lambda^2$ for the soft scales, in accordance with the general discussion in Section~\ref{sec:SCETbasics}.

In our discussion below we will assume that such a regularization is employed. We then introduce the notations\footnote{At leading power in the effective theory, the product $2p_i\cdot p_j$ in the argument of the first logarithm is replaced by $\frac12\,n_i\cdot n_j\,\nb_i\cdot p_i\,\nb_j\cdot p_j$, see (\ref{JS1loop}).}
\begin{equation}
   \beta_{ij} 
   = \ln\frac{-2\sigma_{ij}\,p_i\cdot p_j\,\mu^2}{(-p_i^2)(-p_j^2)}
    \,, \qquad
   L_i = \ln\frac{\mu^2}{-p_i^2}
\end{equation}
for the logarithms of the soft and collinear scales, respectively. The definition of $\beta_{ij}$ generalizes that of the cusp angle in (\ref{cuspangle}) to the case of light-like Wilson lines. The role of the soft scale in (\ref{singlelog}) is played by $\Lambda_s^2=(-p_i^2)(-p_j^2)/(-2\sigma_{ij}\,p_i\cdot p_j)$. Relation (\ref{decomp2}) can now be rewritten as
\begin{equation}\label{rela}
   \beta_{ij} = L_i + L_j - \ln\frac{\mu^2}{-s_{ij}} \,.
\end{equation}

\subsection{General structure of the soft anomalous-dimension matrix}
\label{sec:45}

We are now ready to analyze the structure of the anomalous-dimension matrix of $n$-jet SCET operators. According to the discussion in Section~\ref{sec:decomp}, we can write the decomposition into soft and collinear pieces as
\begin{equation}
   \bm{\Gamma}(\{\underline{p}\},\mu) 
   = \bm{\Gamma}_s(\{\underline{\beta}\},\mu) 
   + \sum_i\,\Gamma_c^i(L_i,\mu) \,,
\end{equation}
where the collinear terms are diagonal in color space. The total anomalous dimension depends on the $n(n-1)/2$ kinematical variables $s_{ij}$, while its soft counterpart depends on the $n(n-1)/2$ cusp angles $\beta_{ij}$, as indicated. The collinear pieces are single-particle terms, each depending on a single collinear scale $L_i$. The general form of the collinear part of the anomalous dimension is known to be of the form \cite{Becher:2003kh}
\begin{equation}\label{gammaci}
   \Gamma_c^i(L_i)
   = - \Gamma_{\rm cusp}^i(\alpha_s)\,L_i + \gamma_c^i(\alpha_s) \,.
\end{equation}
We now substitute for the cusp angles entering the soft anomalous-dimension matrix the expression on the right-hand side of (\ref{rela}). This yields $\bm{\Gamma}_s(\{\underline{s}\},\{\underline{L}\},\mu)$ as a function of the variables $s_{ij}$ and $L_i$. The dependence on the collinear scales must cancel when we combine the soft and collinear contributions to the total anomalous-dimension matrix. We thus obtain the relation
\begin{equation}\label{ourconstraint}
   \frac{\partial\bm{\Gamma}_s(\{\underline{s}\},\{\underline{L}\},
         \mu)}{\partial L_i}
   = \Gamma_{\rm cusp}^i(\alpha_s) \,,
\end{equation}
where the expression on the right-hand side is a unit matrix in color space. This relation provides an important constraint on the momentum and color structures that can appear in the soft anomalous-dimension matrix. A corresponding relation has been derived independently in \cite{Gardi:2009qi}. 

Because the kinematical invariants $s_{ij}$ can be assumed to be linearly independent, relation (\ref{ourconstraint}) implies that $\bm{\Gamma}_s$ depends only linearly on the cusp angles $\beta_{ij}$, see (\ref{rela}). The only exception would be a more complicated dependence on combinations of cusp angles, in which the collinear logarithms cancel. The simplest such combination is
\begin{equation}\label{CCR}
   \beta_{ijkl} 
   = \beta_{ij} + \beta_{kl} - \beta_{ik} - \beta_{jl} 
   = \ln\frac{(-s_{ij})(-s_{kl})}{(-s_{ik})(-s_{jl})} \,,
\end{equation} 
which coincides with the logarithm of the conformal cross ratio $\rho_{ijkl}$ defined in \cite{Gardi:2009qi}. For simplicity, we will use the term ``conformal cross ratio'' in the following also when referring to $\beta_{ijkl}$. This quantity obeys the symmetry properties
\begin{equation}\label{betasym}
   \beta_{ijkl} = \beta_{jilk} = - \beta_{ikjl} = - \beta_{ljki} 
   = \beta_{klij} \,.
\end{equation}
It is easy to show that any combination of cusp angles that is independent of collinear logarithms can be expressed via such cross ratios. Moreover, given four parton momenta, there exist two linearly independent conformal cross ratios, since
\begin{equation}\label{betazerosum}
   \beta_{ijkl} + \beta_{iklj} + \beta_{iljk} = 0 \,,  
\end{equation}
and all other index permutations can be obtained using the symmetry properties in (\ref{betasym}).

Our strategy in Section~\ref{sec:diagrammar} will be to analyze the structure of the soft anomalous-dimension matrix first, since it is constrained by the non-abelian exponentiation theorem and the constraint (\ref{ourconstraint}). The universality of soft gluon interactions implies that the soft contributions only probe the momentum directions and color charges of the external partons, but not their polarization states. Dependence on the parton identities thus only enters via the cusp variables $\beta_{ij}$ and non-trivial color-conserving structures built out of $\bm{T}_i$ generators. If our conjecture (\ref{magic}) is correct, then (\ref{gammaci}) implies that the soft anomalous-dimension matrix should be given by
\begin{equation}\label{Gsoft}
   \bm{\Gamma}_s (\{\underline{\beta}\},\mu)
   = - \sum_{(i,j)}\,\frac{\bm{T}_i\cdot\bm{T}_j}{2}\,
   \gamma_{\rm cusp}(\alpha_s)\,\beta_{ij} 
   + \sum_i \gamma_s^i(\alpha_s) \,,
\end{equation}
where
\begin{equation}
   \gamma^i(\alpha_s) 
   = \gamma_c^i(\alpha_s) + \gamma_s^i(\alpha_s) \,.
\end{equation}
Using relation (\ref{colorrel}) we may indeed confirm that
\begin{equation}\label{tt}
   \frac{\partial\bm{\Gamma}_s}{\partial L_i} 
   = - \sum_{j\ne i}\,\bm{T}_i\cdot\bm{T}_j\,
    \gamma_{\rm cusp}(\alpha_s)
   = C_i\,\gamma_{\rm cusp}(\alpha_s)
   \equiv \Gamma_{\rm cusp}^i(\alpha_s) \,,
\end{equation}
in accordance with the constraint (\ref{ourconstraint}). Note that this result implies Casimir-scaling for the cusp anomalous dimension, since $\Gamma_{\rm cusp}^g(\alpha_s)/\Gamma_{\rm cusp}^q(\alpha_s)=C_A/C_F$. We will come back to the significance of this observation in Section~\ref{sec:Casimir}.

\section{Consistency with collinear limits}
\label{sec:collinear}

Before turning to a diagrammatic study of the anomalous-dimension matrix we discuss one more non-trivial constraint it must obey, which derives from the known behavior of scattering amplitudes in the limit where two or more external partons become collinear. 

In the limit where the momenta of two of the external partons become collinear, an $n$-parton scattering amplitude factorizes into the product of an $(n-1)$-parton scattering amplitude times a universal, process-independent splitting amplitude. This was first shown at tree level in \cite{Berends:1988zn,Mangano:1990by}, and extended to one-loop order in \cite{Bern:1995ix}. An all-order proof was given in \cite{Kosower:1999xi}. Strictly speaking, the proof was constructed for leading-color amplitudes only, but the crucial ingredients are unitarity and analyticity, and it should be possible to extend it to the general case. Collinear factorization holds at the level of the leading singular terms. It is often studied for color-ordered amplitudes, for which the color information is stripped off. The color-stripped splitting amplitudes for the splitting of a parent parton $P$ into collinear partons $a$ and $b$ are usually denoted by $\mbox{Split}_{\sigma_P}(a^{\sigma_a},b^{\sigma_b})$ in the literature, where $\sigma_i$ denote the helicities of the partons. These functions have been calculated at tree level (see, e.g., \cite{Dixon:1996wi}), one-loop order \cite{Kosower:1999rx}, and recently even to two loops \cite{Badger:2004uk}. In contrast, we will study collinear factorization using the color-space formalism, extending the work of \cite{Catani:2003vu} beyond the one-loop approximation. In this framework, the splitting amplitudes are elements of a splitting matrix $\mbox{\bf Sp}(\{p_a,p_b\})$, which acts in the space of color and spin configurations of $(n-1)$-parton scattering amplitudes. As is the case for the scattering amplitudes, the divergence structure of $\mbox{\bf Sp}(\{p_a,p_b\})$ is independent of the spin configuration of the involved partons, and we therefore suppress spin indices in the following. For Catani's formula (\ref{I2}), the consistency with collinear limits was shown in \cite{Bern:2004cz}. 

Consider, for concreteness, the limit where the partons 1 and 2 become collinear and merge into an unresolved parton $P$. We assign momenta 
$p_1=z P$ and $p_2=(1-z) P$ and consider the collinear limit $P^2\to 0$. In this limit the scattering amplitude factorizes in the form
\begin{equation}\label{eq:coll}
   |{\cal M}_{n}(\{p_1,p_2,p_3,\dots,p_n\})\rangle 
   = \mbox{\bf Sp}(\{p_1,p_2\})\,
   |{\cal M}_{n-1}(\{P,p_3,\dots, p_n\})\rangle + \dots \,.
\end{equation}
The matrix of splitting amplitudes encodes the singular behavior of the amplitude $|{\cal M}_n\rangle$ as $p_1||p_2$, and the factorization holds up to terms that are regular in the collinear limit. Analogous relations describe the behavior in limits where more than two partons become collinear. However, it is sufficient for our purposes to focus on the simplest case.

The factorization formula (\ref{eq:coll}) holds both for the dimensionally regularized scattering amplitudes $|{\cal M}_n(\epsilon,\{\underline{p}\})\rangle$ as well as for the minimally subtracted amplitudes $|{\cal M}_n(\{\underline{p}\},\mu)\rangle$ in (\ref{renorm}). Since we know that the divergences of the amplitude can be absorbed into a $\bm{Z}$-factor, equation (\ref{eq:coll}) implies a constraint on the divergences of the splitting amplitudes.
It can be written as
\begin{equation}
   \lim_{\epsilon\to 0} 
   \bm{Z}^{-1}(\epsilon,\{p_1,\dots,p_n\},\mu)\,
   \mbox{\bf Sp}(\epsilon,\{p_1,p_2\})\,
   \bm{Z}(\epsilon,\{P,p_3\dots,p_n\}) 
   = \mbox{\bf Sp}(\{p_1,p_2\},\mu) \,,
\end{equation}
where the matrix of renormalized splitting amplitudes on the right-hand side is finite for $\epsilon\to 0$. From (\ref{Gammadef}) it then follows that this quantity obeys
the RG equation
\begin{equation}
\begin{split}
   \frac{d}{d\ln\mu}\,\mbox{\bf Sp}(\{p_1,p_2\},\mu) 
   &= \bm{\Gamma}(\{p_1,\dots, p_n\},\mu)\,
    \mbox{\bf Sp}(\{p_1,p_2\},\mu) \\
   &\quad\mbox{}- \mbox{\bf Sp}(\{p_1,p_2\},\mu)\,
    \bm{\Gamma}(\{P,p_3\dots,p_n\},\mu) \,.
\end{split}
\end{equation} 
Analogous equations hold for the higher splitting amplitudes $\mbox{\bf Sp}(\{p_1,\dots,p_m\},\mu)$, which describe the limits where more than two partons become collinear. To bring the RG equation into a more useful form, we note that charge conservation implies
\begin{equation}
   (\bm{T}_1 + \bm{T}_2)\,\mbox{\bf Sp}(\{p_1,p_2\},\mu) 
   = \mbox{\bf Sp}(\{p_1,p_2\},\mu)\,\bm{T}_P \,,
\end{equation}
where $\bm{T}_P$ is the color generator associated with the parent parton $P$. Since the matrix of splitting amplitudes commutes with the generators of partons not involved in the splitting process, we can thus commute the anomalous dimension in the second term to the left to obtain
\begin{equation}\label{RGsplit}
   \frac{d}{d\ln\mu}\,\mbox{\bf Sp}(\{p_1,p_2\},\mu)
   = \bm{\Gamma}_{\rm Sp}(\{p_1,p_2\},\mu)\,
   \mbox{\bf Sp}(\{p_1,p_2\},\mu) \,,
\end{equation} 
where we have defined
\begin{equation}\label{Gsplit}
   \bm{\Gamma}_{\rm Sp}(\{p_1,p_2\},\mu) 
   = \bm{\Gamma}(\{p_1,\dots,p_n\},\mu) 
   - \bm{\Gamma}(\{P,p_3\dots,p_n\},\mu) 
   \big|_{\bm{T}_P\to\bm{T}_1+\bm{T}_2} \,.
\end{equation}
The fact that the anomalous dimension of the splitting amplitudes
must be independent of the colors and momenta of the partons not involved in the splitting process, which is a consequence of the factorization formula (\ref{eq:coll}), imposes a non-trivial constraint on the form of the anomalous-dimension matrix. We will explore its implications in Section~\ref{sec:collinearlimit}. 

Assuming the form (\ref{magic}) for the anomalous-dimension matrix $\bm{\Gamma}$, we find that the anomalous dimension of the splitting amplitudes has the simple all-order form
\begin{eqnarray}\label{gammaSp}
   \bm{\Gamma}_{\rm Sp}(\{p_1,p_2\},\mu)
   &=& \gamma_{\rm cusp} \left[ 
    \bm{T}_1\cdot\bm{T}_2\,\ln\frac{\mu^2}{-s_{12}} 
    + \bm{T}_1\cdot (\bm{T}_1+\bm{T}_2)\,\ln z 
    + \bm{T}_2\cdot (\bm{T}_1+\bm{T}_2)\,\ln(1-z) \right] \nonumber\\
   &&\mbox{}+ \gamma^1 + \gamma^2 - \gamma^P \,,
\end{eqnarray}
where $\gamma^P$ is the anomalous dimension associated with the unresolved parton $P$. Note that the momentum-dependent terms in the result are insensitive to the nature of the partons involved in the splitting process. The divergent part of the one-loop splitting amplitudes for $m$ partons in the color-space formalism was given in \cite{Catani:2003vu}. Expanding the result obtained there for the case $m=2$, we find
\begin{equation}
\begin{split}
   \mbox{\bf Sp}_{\rm 1-loop}(\epsilon,\{p_1,p_2\}) 
   &= \frac{\alpha_s}{4\pi}\,\bigg[ \left( \frac{2}{\epsilon^2} 
     +  \frac{2}{\epsilon}\,\ln\frac{\mu^2}{-s_{12}} \right)
    \bm{T}_1\cdot \bm{T}_2 \\
   &\quad\mbox{} + \frac{2}{\epsilon}\,\Big[ 
    \bm{T}_1\cdot (\bm{T}_1+\bm{T}_2)\,\ln z 
    + \bm{T}_2\cdot  (\bm{T}_1+\bm{T}_2)\,\ln(1-z) \Big] 
    \\
   &\quad\mbox{} + \frac{1}{2\epsilon}
    \left( \gamma_0^1 + \gamma_0^2 - \gamma_0^a \right) 
    + {\cal O}(\epsilon^0) \bigg]\,
    \mbox{\bf Sp}_{\rm tree}(\{p_1,p_2\}) \,,
\end{split}
\end{equation}
which is in agreement with the result obtained by solving the RG equation (\ref{RGsplit}). Beyond one-loop order the splitting amplitudes are given by rather complicated expressions \cite{Badger:2004uk}; however, we have checked that their IR divergences can indeed be obtained from the simple anomalous dimension in (\ref{gammaSp}), which only contains single logarithms of the momentum fractions $z$ and $(1-z)$. The exact form of the anomalous-dimension matrix in (\ref{gammaSp}) is an important by-product of our analysis.

\section{Diagrammatic analysis}
\label{sec:diagrammar}

We now present a detailed diagrammatic study of the general structure of the soft anomalous-dimension matrix up to three-loop order, implementing the constraints that follow from the non-abelian exponentiation theorem and RG invariance of the effective theory. At two-loop order we will recover the form found in \cite{MertAybat:2006wq,MertAybat:2006mz} from a simple symmetry argument. In these papers only the cusp piece of the soft anomalous-dimension matrix was studied, which is legitimate given that the non-logarithmic terms can be shown to be diagonal in color space. We find that this property is no longer trivial beyond two-loop order. 

\begin{figure}
\begin{center}
\includegraphics[width=10.7cm]{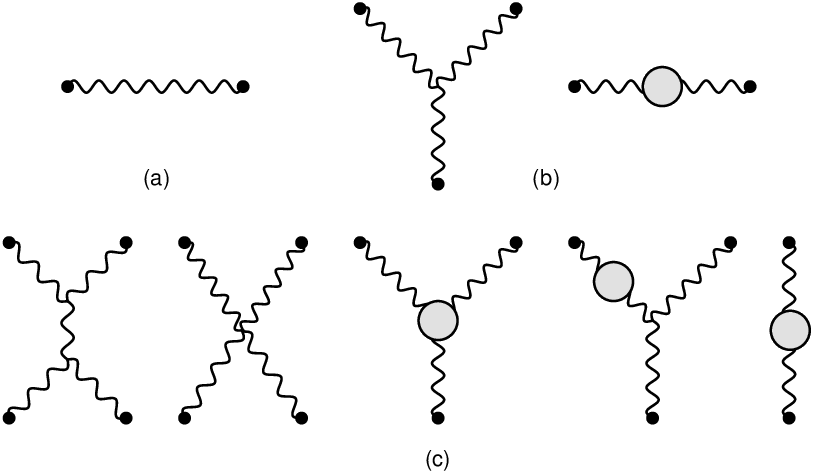}
\end{center}
\vspace{-4mm}
\caption{\label{fig:webs}
One-loop (a), two-loop (b), and three-loop (c) connected webs contributing to the soft anomalous-dimension matrix. The dots represent color generators, which appear when the gluons are attached to Wilson lines. In each set, only the first web gives rise to a new color structure.}
\end{figure}

The non-abelian exponentiation theorem restricts the color structures that can potentially appear in the soft anomalous-dimension matrix. They are obtained by considering single connected webs, whose ends can be attached in arbitrary ways to the $n$ Wilson lines in the soft operator in (\ref{Slines}). In general, single connected webs at $L$-loop order carry between 2 and $(L+1)$ color generators $\bm{T}$. In Figure~\ref{fig:webs} we show the webs appearing up to three-loop order. The dashed blobs represent self-energy or vertex functions, which have color structure $\delta^{ab}$ and $-if^{abc}$. The color structures of the three- and four-gluon vertices can be expressed in terms of $f^{abc}$ symbols.

In our analysis in this section we only use basic properties of the Lie algebra of the gauge group, which can be summarized as
\begin{equation}\label{Liegroup}
\begin{split}
   [\bm{T}^a,\bm{T}^b] &= if^{abc}\,\bm{T}^c \,, \qquad
    f^{abc} f^{abd} = C_A\,\delta^{cd} \,, \\
   \mbox{tr}_{\rm adj.} \left( \bm{T}^a\,\bm{T}^b\,\bm{T}^c \right)
   &= i f^{ade} f^{beg} f^{cgd} = \frac{i C_A}{2}\,f^{abc} \,.
\end{split}
\end{equation}
The last relation follows from the Jacobi identity, i.e., the first relation in the adjoint representation. While our explicit analysis refers to $SU(N)$ non-abelian gauge theories, its validity extends to other gauge groups as well.

\subsection{One-loop analysis}

In this case the relevant web consists of a single gluon, as shown in Figure~\ref{fig:webs}(a). If it is attached to two different Wilson lines $i$ and $j$, then the resulting color structure is $\bm{T}_i\cdot\bm{T}_j$. In this case non-trivial momentum dependence can arise, which can lead to a factor $\beta_{ij}$. Recall that only linear dependence on the cusp angle is allowed. For terms without momentum dependence, the sum over parton legs reduces the color structure to a diagonal one, since relation (\ref{colorrel}) can be applied in this case. Likewise, if the ends of the exchanged gluon are attached to a single Wilson line $i$, then the color structure is $\bm{T}_i^2=C_i$. It follows that at one-loop order the soft anomalous-dimension matrix is indeed of the form (\ref{Gsoft}). 

\subsection{Two-loop analysis}
\label{sec:2loop}

In this case two webs need to be considered, which are depicted in Figure~\ref{fig:webs}(b). The connected web containing the gluon self-energy has the same color structure as a single gluon exchange, and hence it does not lead to any new structures in the result (\ref{Gsoft}). The color structure of the three-gluon web is proportional to $-if^{abc}$ times three color generators, one for each leg. There are thus three possibilities, which we consider separately. 

If all gluons are attached to a single Wilson line, then the resulting color structure is
\begin{equation}
   -if^{abc}\,\bm{T}_i^a\,\bm{T}_i^b\,\bm{T}_i^c
   = \frac{C_A C_i}{2} \,.
\end{equation}
In this case no momentum dependence can arise. If the gluons are attached to two different Wilson lines $i$ and $j$, then the resulting color structure is (recall that generators belonging to different partons commute)
\begin{equation}
   -if^{abc}\,\bm{T}_i^a\,\bm{T}_i^b\,\bm{T}_j^c
   = \frac{C_A}{2}\,\bm{T}_i\cdot\bm{T}_j \,.
\end{equation}
In this case momentum dependence can arise, since two partons are involved in the loop diagram. It is thus possible to get a factor $\beta_{ij}$ or a constant. In any case we obtain the same structures as at one-loop order. Finally, if the gluons are attached to three different Wilson lines $i$, $j$, and $k$, then the resulting color structure
\begin{equation}
   -if^{abc}\,\bm{T}_i^a\,\bm{T}_j^b\,\bm{T}_k^c
\end{equation}
is totally anti-symmetric in the parton indices, and it would therefore need to multiply a totally anti-symmetric momentum-dependent structure formed out of the three kinematical invariants $\beta_{ij}$, $\beta_{jk}$, and $\beta_{ki}$. However, no such structure exists that would be consistent with our constraint (\ref{ourconstraint}), since it requires linearity in the cusp angles in cases where less than four parton momenta are involved. A nonlinear structure such as 
\begin{equation}
\begin{split}
   & (\beta_{ij}-\beta_{jk})(\beta_{jk}-\beta_{ki})
    (\beta_{ki}-\beta_{ij}) \\
   &= \left( \ln\frac{-s_{ij}}{-s_{jk}} + L_i - L_k \right)
    \left( \ln\frac{-s_{jk}}{-s_{ki}} + L_j - L_i \right)
    \left( \ln\frac{-s_{ki}}{-s_{ij}} + L_k - L_j \right) ,
\end{split}
\end{equation}
which is reminiscent of that appearing in (\ref{H2}), cannot be written as a sum of hard and collinear contributions. Recall that this structure arises in Catani's subtraction operator $\bm{I}^{(2)}$ only because in \cite{Catani:1998bh} the subtraction of IR poles is not implemented in a minimal scheme, giving rise to a term sensitive to the finite parts of the hard scattering amplitudes. 

This simple symmetry argument explains the cancellations observed in \cite{MertAybat:2006wq,MertAybat:2006mz}, where two-loop diagrams with gluon attachments to three different parton legs were shown to vanish. It follows that at two-loop order the soft anomalous-dimension matrix is still of the form (\ref{Gsoft}).

\subsection{Three-loop analysis}
\label{sec:3loops}

The single connected webs in Figure~\ref{fig:webs}(c) containing insertions of self-energy or vertex functions have the same color structure as the corresponding two-loop webs and hence give nothing new compared with the discussion at two-loop order. This explains the cancellations observed in \cite{Dixon:2009gx}, where three-loop diagrams with fermionic self-energy insertions and gluon attachments to three different parton legs were shown to vanish. It thus suffices to consider the two four-gluon webs, both of which have the color structure $f^{ade} f^{bce}\,\bm{T}_i^a\,\bm{T}_j^b\,\bm{T}_k^c\,\bm{T}_l^d$. If two or more color generators act on the same parton (i.e., if two or more of the indices $i,j,k,l$ coincide), then the products of generators with the same parton index can be decomposed into symmetric and anti-symmetric products. Using the Lie-algebra relations in (\ref{Liegroup}), the anti-symmetric products can always be reduced to structures containing fewer generators. In this case we obtain one of the color structures already present in the two-loop case. It is therefore sufficient to consider only the symmetric product of four color generators, as defined in (\ref{symprod}). We thus introduce the notation
\begin{equation}\label{Tdef}
   {\cal T}_{ijkl} = f^{ade} f^{bce}
   \left( \bm{T}_i^a\,\bm{T}_j^b\,\bm{T}_k^c\,\bm{T}_l^d \right)_+ .
\end{equation}
This object has the symmetry properties
\begin{equation}\label{Tsym}
   {\cal T}_{ijkl} = {\cal T}_{jilk} 
   = - {\cal T}_{ikjl} = - {\cal T}_{ljki} 
   = {\cal T}_{klij} \,.
\end{equation}
If all four parton indices are different, then these are precisely the symmetry properties of the conformal cross ratio $\beta_{ijkl}$ in (\ref{CCR}). Note that if four or three indices coincide, then the symmetric product (\ref{Tdef}) vanishes. If two indices coincide, the non-vanishing index combinations are
\begin{equation}\label{Tijklcombs}
\begin{aligned}
   {\cal T}_{iijj} &= - {\cal T}_{ijij}
    = f^{ade} f^{bce} \left( \bm{T}_i^a\,\bm{T}_i^b \right)_+ 
    \left( \bm{T}_j^c\,\bm{T}_j^d \right)_+ \,, \\
   {\cal T}_{iijk} &= - {\cal T}_{ijik} 
    = - {\cal T}_{jiki} = {\cal T}_{jkii}
    = f^{ade} f^{bce} \left( \bm{T}_i^a\,\bm{T}_i^b \right)_+
    \bm{T}_j^c\,\bm{T}_k^d \,.
\end{aligned}
\end{equation}

The four gluons of the connected webs shown in the first two graphs in Figure~\ref{fig:webs}(c) can be attached to up to four Wilson lines. There are thus several cases that need to be distinguished. If the gluons are attached to two different Wilson lines $i$ and $j$, then the color structure shown in the first line of (\ref{Tijklcombs}) can arise. Since two parton legs are involved in the loop diagrams, a dependence on $\beta_{ij}$ can arise, which is at most linear. We thus have the possibilities $\beta_{ij}$ or a constant. If the four gluons are attached to three different Wilson lines $i$, $j$, and $k$, then the color structure shown in the second line of (\ref{Tijklcombs}) can arise, which is symmetric in $j$ and $k$. It follows that this structure can be combined with a symmetric combination of the variables $\beta_{ij}$, $\beta_{jk}$, and $\beta_{kl}$. This leaves the three possibilities $\beta_{jk}$, $(\beta_{ij}+\beta_{ik})$, and a constant. Finally, if the four gluons are attached to four different Wilson lines $i$, $j$, $k$, and $l$, then the color structure ${\cal T}_{ijkl}$ must be combined with a momentum structure with the same symmetry properties as those shown in (\ref{Tsym}). In this case four parton legs are involved in the loop integration, and hence the result can depend on the six cusp angles that can be formed out of the four parton momenta. However, the anti-symmetry of ${\cal T}_{ijkl}$ in $(j,k)$ and $(i,l)$ eliminates $\beta_{jk}$ and $\beta_{il}$. Indeed, starting with any linear function of the cusp angles, symmetry arguments can be used to replace
\begin{equation}\label{4sym}
\begin{aligned}
   \sum_{(i,j,k,l)}\!{\cal T}_{ijkl}\,\beta_{ij} 
   &= \sum_{(i,j,k,l)}\!{\cal T}_{ijkl}\,\beta_{kl}
    = \sum_{(i,j,k,l)}\!{\cal T}_{ijkl}\,
    \frac{\beta_{ij}+\beta_{kl}-\beta_{ik}-\beta_{jl}}{4}
    = \sum_{(i,j,k,l)}\!{\cal T}_{ijkl}\,\frac{\beta_{ijkl}}{4} 
    \,, \\
   \sum_{(i,j,k,l)}\!{\cal T}_{ijkl}\,\beta_{ik} 
   &= \sum_{(i,j,k,l)}\!{\cal T}_{ijkl}\,\beta_{jl} 
    = \sum_{(i,j,k,l)}\!{\cal T}_{ijkl}\,
    \frac{\beta_{ik}+\beta_{jl}-\beta_{ij}-\beta_{kl}}{4}
    = - \sum_{(i,j,k,l)}\!{\cal T}_{ijkl}\,\frac{\beta_{ijkl}}{4} 
    \,, \\
   \sum_{(i,j,k,l)}\!{\cal T}_{ijkl}\,\beta_{jk} 
   &= \sum_{(i,j,k,l)}\!{\cal T}_{ijkl}\,\beta_{il}
    = 0 \,.
\end{aligned}
\end{equation}
This leaves us with the structure ${\cal T}_{ijkl}\,\beta_{ijkl}$. 

Since conformal cross ratios are invariant under the transformation from soft to hard scales, the factorization constraint (\ref{ourconstraint}) does not prevent us from considering more complicated functions of such ratios. The most general possibility would be to allow a function depending on two linearly independent combinations of conformal cross ratios. Using (\ref{betazerosum}), we can write
\begin{equation}
   F(\beta_{ijkl},\beta_{iklj}-\beta_{iljk}) \,,  
\end{equation}
where the second argument is invariant under all of the index permutations in (\ref{Tsym}). As long as $F$ is odd in its first argument, $F(x,y)=-F(-x,y)$, this ansatz respects the symmetry properties of the color structure ${\cal T}_{ijkl}$. It is not easy to see how a non-trivial function of conformal cross ratios could arise from a calculation of Feynman diagrams. According to (\ref{lnZ}), any new structure in the anomalous dimension  can first arise in the three-loop coefficients $\bm{\Gamma}_2$ and $\Gamma'_2$. Since the relevant Feynman diagrams are free of subdivergences, they do not generate higher poles than $1/\epsilon^2$. One then expects at most single logarithms to appear in the coefficient of the $1/\epsilon$ pole term. Indeed, the structures in (\ref{4sym}) do not contain a cusp logarithm and therefore can only contribute to the $\bm{\Gamma}_2/\epsilon$ term in (\ref{lnZ}). An example of a structure that can give rise to such a term is
\begin{equation}
   \frac{1}{\epsilon^2} \left[ 
   \left( \frac{\mu^2}{-s_{ij}} \right)^{3\epsilon}
   + \left( \frac{\mu^2}{-s_{kl}} \right)^{3\epsilon}
   - \left( \frac{\mu^2}{-s_{ik}} \right)^{3\epsilon}
   - \left( \frac{\mu^2}{-s_{jl}} \right)^{3\epsilon} \right]
   = - \frac{3}{\epsilon}\,\beta_{ijkl} + \dots \,.
\end{equation}
We thus consider a linear dependence on the conformal cross ratio $\beta_{ijkl}$ as the most plausible possibility. However, in our discussion below we will allow for an arbitrary dependence consistent with the symmetries of the problem.

At this point we have exhausted the new structures that could in principle contribute to the soft anomalous-dimension matrix at three-loop order. Absorbing all terms not fitting the simple forms shown in (\ref{Gsoft}) into a quantity $\bm{\Delta\Gamma}_s$, we obtain 
\begin{equation}\label{Deltags}
\begin{split}
   \bm{\Delta\Gamma}_s
   &= \sum_{(i,j)}\,{\cal T}_{iijj}\,
    \Big[ f_1(\alpha_s)\,\beta_{ij} + f_2(\alpha_s) \Big] \\
   &\quad\mbox{}+ \sum_{(i,j,k)} {\cal T}_{iijk}\,
    \Big[ f_3(\alpha_s)\,\beta_{jk} 
    + f_4(\alpha_s) (\beta_{ij} + \beta_{ik}) 
    + f_5(\alpha_s) \Big] \\
   &\quad\mbox{}+ \!\sum_{(i,j,k,l)}\!
    {\cal T}_{ijkl}\,\Big[ f_6(\alpha_s)\,\beta_{ijkl}
    + F(\beta_{ijkl},\beta_{iklj}-\beta_{iljk}) \Big] \,.
\end{split}
\end{equation}
The functions $f_i(\alpha_s)$ and $F(x,y)$, which represents a possible dependence on conformal cross ratios that is more complicated than a linear $\beta_{ijkl}$ term, start at three-loop order. We suppress the argument $\alpha_s$ in the latter function for brevity. Expression (\ref{Deltags}) can be simplified considerably by performing the sums over those parton indices in the second and third lines not involved in the various $\beta_{ij}$ factors. In particular, the four-parton term linear in $\beta_{ijkl}$ can be simplified by using one of the first two relations in (\ref{4sym}) and performing the sums over the two free indices. Recall that generators belonging to different partons commute, so a generator whose index is summed over can always be moved to the right of all other generators, and then relation (\ref{singlet}) can be applied. After a straightforward calculation, we obtain
\begin{equation}\label{eq63}
\begin{split}
   \bm{\Delta\Gamma}_s
   &~\widehat=~ \sum_{(i,j)}\,{\cal T}_{iijj}
    \left[ \Big( f_1(\alpha_s) - 2 f_4(\alpha_s) 
     - 4 f_6(\alpha_s) \Big)\,\beta_{ij} 
     + f_2(\alpha_s) - f_5(\alpha_s) \right] \\
   &\quad\mbox{}+ \sum_{(i,j,k)} {\cal T}_{iijk}\,
    \Big[ f_3(\alpha_s) - 4 f_6(\alpha_s) \Big]\,\beta_{jk} 
    + \!\sum_{(i,j,k,l)}\!{\cal T}_{ijkl}\,
    F(\beta_{ijkl},\beta_{iklj}-\beta_{iljk}) \,,
\end{split}
\end{equation}
where the symbol ``$\widehat=$'' means that the two expressions agree up to trivial color structures, which can be absorbed into (\ref{Gsoft}) and hence can be dropped from $\bm{\Delta\Gamma}_s$. Next, we evaluate the constraint (\ref{ourconstraint}) for the above expression, which implies
\begin{equation}\label{test}
\begin{split}
   \frac{\partial\bm{\Delta\Gamma}_s}{\partial L_i}
   &= 2\,\sum_{j\ne i}\,{\cal T}_{iijj}\,
    \Big[ f_1(\alpha_s) - 2 f_4(\alpha_s) - 4 f_6(\alpha_s) \Big] 
    + 2\!\sum_{(j\ne i,k\ne i)}\!{\cal T}_{jjik}\,
    \Big[ f_3(\alpha_s) - 4 f_6(\alpha_s) \Big] \\
   &= 2\,\sum_{j\ne i}\,{\cal T}_{iijj}\,
    \Big[ f_1(\alpha_s) - f_3(\alpha_s) - 2 f_4(\alpha_s) \Big] 
    + \frac{C_A^2 C_i}{4}\,\Big[ f_3(\alpha_s)
    - 4 f_6(\alpha_s) \Big] \,,
\end{split}
\end{equation}
where in the last step we have performed the sum over $k$. Relation (\ref{ourconstraint}) requires that this result be proportional to the unit matrix in color space times a coefficient depending only on the representation of parton $i$, which is not satisfied for the color structure of the first term in the last equation. To see this, suppose there exists a constant $k_i$ depending only on the representation of parton $i$, such that $\sum_{j\ne i} {\cal T}_{iijj}=k_i\,\bm{1}$. Taking traces over the color indices of either parton $i$, or of all other partons, we find that $k_i$ would have to be proportional to $\sum_{j\ne i} C_j$, which evidently is not independent of the color representations of the remaining partons involved in the scattering process. We must therefore require that
\begin{equation}\label{rela1}
   f_3(\alpha_s) = f_1(\alpha_s) - 2f_4(\alpha_s) \,.
\end{equation}
The most general expression for the extra terms can thus be written as (with obvious redefinitions of the coefficient functions)
\begin{equation}\label{DGsfinal}
   \bm{\Delta\Gamma}_s
   = \sum_{(i,j)}\,{\cal T}_{iijj}\,
   \Big[ \bar f_1(\alpha_s)\,\beta_{ij} + \bar f_2(\alpha_s) \Big] 
   + \sum_{(i,j,k)} {\cal T}_{iijk}\,\bar f_1(\alpha_s)\,\beta_{jk} 
   + \!\sum_{(i,j,k,l)}\!{\cal T}_{ijkl}\,
   F(\beta_{ijkl},\beta_{iklj}-\beta_{iljk}) \,.
\end{equation}
It follows that using arguments based on factorization and non-abelian exponentiation alone, one cannot exclude color and momentum structures in the soft anomalous-dimension matrix that are more complicated that those in (\ref{Gsoft}).

Inverting the relations between color structures that led to (\ref{eq63}) and expressing the result in terms of structures containing maximal numbers of color generators, we find that the most general form of the additional contributions to the anomalous-dimension matrix $\bm{\Gamma}$ of $n$-jet SCET operators in (\ref{magic}) is 
\begin{equation}\label{DG3final}
\begin{split}
   \bm{\Delta\Gamma}_3(\{\underline{p}\},\mu)
   &= - \frac{\bar f_1(\alpha_s)}{4} \sum_{(i,j,k,l)}\!
    {\cal T}_{ijkl}\,\ln\frac{(-s_{ij})(-s_{kl})}{(-s_{ik})(-s_{jl})}
    - \bar f_2(\alpha_s) \sum_{(i,j,k)} {\cal T}_{iijk} \\
   &\quad\mbox{}+ \!\sum_{(i,j,k,l)}\!{\cal T}_{ijkl}\,
    F(\beta_{ijkl},\beta_{iklj}-\beta_{iljk}) \,,
\end{split}
\end{equation}
where the subscript ``3'' indicates that these structures could first arise at three-loop order. At this order the entire contribution $\bm{\Delta\Gamma}_3$ is proportional $\alpha_s^3$ and all color dependence is explicit, i.e.\ it arises only from the tensors ${\cal T}_{ijkl}$. To determine the coefficient function $F(x,y)=-F(-x,y)$ and the numerical coefficients multiplying the other two terms it would suffice to calculate an arbitrary four-parton amplitude at three-loop order. If any one of the terms in (\ref{DG3final}) did not vanish, then our conjecture (\ref{magic}) for the structure of the anomalous-dimension matrix would have to be modified starting at three-loop order. In Sections~\ref{sec:simplifications} and \ref{sec:collinearlimit} we will show that the coefficients $\bar f_1(\alpha_s)$ and $\bar f_2(\alpha_s)$ indeed vanish to all orders in perturbation theory, and that the function $F(x,y)$ must vanish in all two-parton collinear limits, which is compatible with it being zero for all values of its arguments. We note, however, that even if the contribution proportional to $F(x,y)$ would not turn out to be zero no explicit $\mu$ dependence enters in (\ref{DG3final}), so that
\begin{equation}
   \frac{\partial}{\partial\ln\mu}\,
   \bm{\Delta\Gamma}_3(\{\underline{p}\},\mu) = 0 \,,
\end{equation}
and hence there would not be a contribution of this structure to the function $\Gamma'(\alpha_s)$ in (\ref{Gampr}). It follows that in (\ref{lnZ}) a modification could first enter in the $\bm{\Gamma}_2/\epsilon$ term at three-loop order. Equivalently, the structure of the cusp logarithms in the anomalous-dimension matrix remains unaffected up to and including three loops, while the non-cusp terms remain unaffected at least to two-loop order. Based on our result (\ref{magic}), and irrespective of whether the additional terms in (\ref{DG3final}) vanish or not, it is therefore possible to resum large Sudakov logarithms at next-to-next-to-leading-logarithmic accuracy. This is sufficient for most practical applications, since it allows the resummation of all Sudakov logarithms which appear in next-to-leading order calculations of $n$-jet processes. Beyond this accuracy, two-loop calculations of amplitudes with $n$ partons are required to obtain the necessary matching coefficients.

\subsection{Higher Casimir contributions to the cusp anomalous dimension}
\label{sec:Casimir}

For the special case of two-jet operators, the simple form (\ref{magic}) implies {\em Casimir-scaling\/} of the cusp anomalous dimension, i.e., the cusp anomalous dimensions of quarks and gluons are related to each other by the ratio of the eigenvalues $C_i$ of the quadratic Casimir operators:
\begin{equation}
   \frac{\Gamma_{\rm cusp}^q(\alpha_s)}{C_F}
   = \frac{\Gamma_{\rm cusp}^g(\alpha_s)}{C_A}
   = \gamma_{\rm cusp}(\alpha_s) \,,
\end{equation}
see (\ref{2jet}) and (\ref{tt}). This relation is indeed satisfied at three-loop order \cite{Moch:2004pa}. To this order Casimir scaling is a consequence of non-abelian exponentiation, as can be seen from our analysis above: restricted to the two-jet case, all possible color structures arising up to three-loop order are proportional to $C_i$. Beyond three loops non-abelian exponentiation no longer automatically implies Casimir scaling \cite{Frenkel:1984pz}, and there are arguments based on calculations using the AdS/CFT correspondence \cite{Maldacena:1997re,Gubser:1998bc,Witten:1998qj} suggesting a violation at higher orders \cite{Armoni:2006ux,Alday:2007hr,Alday:2007mf}. The new color structures would involve higher Casimir invariants such as those appearing in the four-loop $\beta$-function of non-abelian gauge theories \cite{vanRitbergen:1997va,Czakon:2004bu}.

For the case of ${\cal N}=4$ SYM in the strong coupling limit, $\lambda=g_s^2 N_c\to\infty$, a violation of Casimir scaling was found in \cite{Armoni:2006ux} by considering a Wilson loop in a $k$-dimensional antisymmetric representation of $SU(N_c)$ in the limit where $N_c$ and $k$ go to infinity at fixed ratio $N_c/k$. Since the calculation was performed in the strong-coupling limit, it does not predict if and at which order in the weak-coupling expansion the effect would appear. On the other hand, it is not implausible that it might appear at some order in perturbation theory, since the perturbative resummation of ladder diagrams contributing to Wilson loops in ${\cal N}=4$ SYM indeed produces, when reexpanded for large $\lambda$, the $e^{\sqrt{\lambda}}$ behavior characteristic for the strong-coupling limit \cite{Erickson:2000af}. Also, in \cite{Eden:2006rx,Beisert:2006ez} an all-order form of the cusp anomalous dimension of planar ${\cal N}=4$ SYM was proposed, which is given by the solution of a certain integral equation. This conjecture has been checked by four-loop calculations in the weak-coupling limit \cite{Bern:2005iz} and to second order in the strong-coupling expansion using AdS/CFT and a two-loop superstring calculation \cite{Roiban:2007dq}. 

Higher Casimir invariants can be constructed by considering symmetrized traces
\begin{equation}
   d_R^{a_1 a_2\dots a_n} 
   = \mbox{tr} \big[ \left( \bm{T}_R^{a_1}\,\bm{T}_R^{a_2}\ldots\,
   \bm{T}_R^{a_n} \right)_+ \big]
\end{equation} 
of generators in a representation $R$. Any such trace contracted with $n$ generators defines a Casimir invariant, since
\begin{equation}
   C_n(R,R') = d_R^{a_1 a_2\dots a_n}\,
   \bm{T}_{R'}^{a_1}\,\bm{T}_{R'}^{a_2}\ldots\,\bm{T}_{R'}^{a_n}
\end{equation} 
commutes with all generators in the representation $R$ of the group. If $R$ is irreducible, then Schur's lemma implies that $C_n(R,R')$ is proportional to the unit matrix. These Casimir invariants are, however, not all independent. To obtain an independent set of Casimir operators it is sufficient to consider symmetric traces in the fundamental representation to define the $d$-symbols, since $d_R^{a_1 a_2\dots a_n}=I_n(R)\,d_F^{a_1 a_2\dots a_n}$ with a representation-dependent index $I_n(R)$. Furthermore, the invariants can be redefined,  $d^{a_1\dots a_n}\to d_\perp^{a_1\dots a_n}$, such that they fulfill the orthogonality conditions $d_\perp^{a_1\dots a_l\dots a_n}\,d_\perp^{a_1\dots a_l}=0$ \cite{Okubo:1981td,Okubo:1982dt}. For $SU(N)$ groups, $N-1$ independent invariants can be constructed in this way. More details on the evaluation of group-theory factors appearing in Feynman diagrams can be found in \cite{vanRitbergen:1998pn}.

\begin{figure}
\begin{center}
\includegraphics[height=2.5cm]{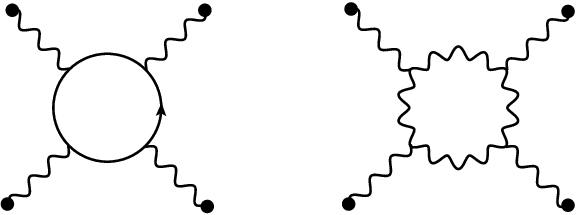}
\end{center}
\vspace{-4mm}
\caption{\label{fig:highercasimir}
Four-loop connected webs involving higher Casimir invariants.}
\end{figure}

Let us now consider possible contributions of these new color structures to the cusp part of the soft anomalous-dimension matrix. The case $n=3$ is irrelevant. The corresponding connected web,  depicted in the middle graph in Figure~\ref{fig:webs}(c), consists of three gluons attached to a gluon or fermion loop. These contributions have antisymmetric color structure $f_{abc}$. Symmetric traces of four color generators do arise, however, from the diagrams shown in Figure~\ref{fig:highercasimir}. The corresponding single connected webs can contribute to the soft anomalous-dimension matrix starting at four-loop order. Our goal is to study the most general contributions of these webs proportional to a cusp logarithm. A complete classification of potential new color and momentum structures that could arise at four-loop order is left for future work.

Using the notation
\begin{equation}
   {\cal D}_{ijkl} 
   = d_F^{abcd}\,\bm{T}_i^a\,\bm{T}_j^b\,\bm{T}_k^c\,\bm{T}_l^d 
   = d_F^{abcd} \left( \bm{T}_i^a\,\bm{T}_j^b\,\bm{T}_k^c\,
    \bm{T}_l^d \right)_+ ,
\end{equation}
the possible contributions to the soft anomalous-dimension matrix linear in cusp angles have the following structures: $\beta_{ij}\,{\cal D}_{iijj}$ and $\beta_{ij}\,{\cal D}_{iiij}$ (gluon attachments to two different Wilson lines), $\beta_{jk}\,{\cal D}_{iijk}$ and $(\beta_{ij}+\beta_{ik})\,{\cal D}_{iijk}$ (attachments to three different Wilson lines), or $\beta_{ij}\,{\cal D}_{ijkl}$ (attachments to four different Wilson lines). Here we have exploited the fact that ${\cal D}_{ijkl}$ is totally symmetric in its indices. Using color conservation to evaluate the sums over free parton indices, the result can be reduced to
\begin{equation}\label{eq:hc1}
   \bm{\Delta\Gamma}_s^{\rm cusp}
   = \sum_{(i,j)}\,\beta_{ij}\,\Big[ {\cal D}_{iijj}\,g_1(\alpha_s) 
    + {\cal D}_{iiij}\,g_2(\alpha_s) \Big] 
   + \sum_{(i,j,k)} \beta_{ij}\,{\cal D}_{ijkk}\,g_3(\alpha_s) \,,
\end{equation}
where the superscript ``cusp'' indicates that we only focus on new structures linear in cusp angles. The coefficient functions contain in general two terms of the form $g_i(\alpha_s)=n_f\,g_i^F(\alpha_s)+I_4(A)\,g_i^A(\alpha_s)$, see Figure~\ref{fig:highercasimir}. They start at ${\cal O}(\alpha_s^4)$.

Let us now evaluate the condition (\ref{ourconstraint}), which implies
\begin{equation}
   \frac{\partial\bm{\Delta\Gamma}_s^{\rm cusp}}{\partial L_i}
   = - C_4(F,R_i)\,g_2(\alpha_s)    
   + \sum_{j\ne i} \left[ 2 {\cal D}_{iijj} 
   \Big( g_1(\alpha_s) - g_3(\alpha_s) \Big)
   + {\cal D}_{ijjj} \Big( g_2(\alpha_s) - 2g_3(\alpha_s) \Big)  
   \right] .
  \end{equation}  
Only the first term on the right-hand side is of the required form and can be absorbed into the jet-function anomalous dimension, so that the factorization constraint (\ref{ourconstraint}) implies
\begin{equation}\label{g3rela}
   g_3(\alpha_s) = g_1(\alpha_s) = \frac{g_2(\alpha_s)}{2} \,.
\end{equation}
The higher-Casimir cusp terms must thus have the form
\begin{equation}\label{nonCasimirstr}
   \bm{\Delta\Gamma}_s^{\rm cusp}
   = g_1(\alpha_s)\,\bigg[ \sum_{(i,j)}\,\beta_{ij}\, 
    \big( {\cal D}_{iijj} + 2 {\cal D}_{iiij} \big)
   + \sum_{(i,j,k)} \beta_{ij}\,{\cal D}_{ijkk} \bigg] \,.
\end{equation}
It is remarkable that the factorization constraint determines the structure of this term uniquely up to an overall coefficient function.

The corresponding contribution to the four-loop anomalous-dimension matrix of $n$-jet SCET operators is given by
\begin{equation}\label{DG4cusp}
   \bm{\Delta\Gamma}_4^{\rm cusp}
   = - g_1(\alpha_s)\,\bigg[ \sum_{(i,j)}\,\ln\frac{\mu^2}{-s_{ij}}\, 
   \big( {\cal D}_{iijj} + 2 {\cal D}_{iiij} \big)
   + \sum_{(i,j,k)} \ln\frac{\mu^2}{-s_{ij}}\,{\cal D}_{ijkk} \bigg]
   \,.
\end{equation}
According to (\ref{tt}), the cusp anomalous dimension for two-jet operators now receives a contribution not proportional to the quadratic Casimir operator of the gauge group. It is given by
\begin{equation}\label{nonCasimir}
   \Gamma_{\rm cusp}^i(\alpha_s) 
   = C_i\,\gamma_{\rm cusp}(\alpha_s) - 2 g_1(\alpha_s)\,C_4(F,R_i)
   \,.
\end{equation}
The four-loop cusp anomalous dimension is known for ${\cal N}=4$ SYM in the planar limit \cite{Bern:2006ew}. However, we show in Appendix~B that the higher Casimir contributions, as well as the three-loop structures (\ref{DG3final}) discussed above, are subleading in the $N_c\to\infty$ limit. These structures are therefore not visible in the planar limit.

\subsection{A symmetry argument}
\label{sec:simplifications}

Below, we will show that (with one possible exception) the additional structures considered in Section~\ref{sec:3loops} and \ref{sec:Casimir} can be excluded, because they do not have the correct properties in the two-parton collinear limit. Before turning to this discussion, however, we find it instructive to present a physical argument suggesting that the additional structures should be absent. To this end, note that the simple form of color-symmetrized soft gluon attachments to a set of Wilson lines discussed in Section~\ref{sec:softattachments} implies restrictions on the form of the various terms in (\ref{Deltags}). The Feynman rules discussed there imply that, irrespective of which Wilson lines the gluons attach to, one gets the same loop integral apart from substitutions of $n_i$ vectors. 

If the symmetry properties of the diagrams also hold for the anomalous dimensions, they imply that the extra structures vanish due to color conservation, as we will show below. What makes the argument somewhat subtle is that even for IR-finite quantities, individual Feynman diagrams can contain IR divergences. These can manifest themselves in scaleless integrals, for which the expansion around $d=4$ does not commute with the use of symmetry relations. For example, if one considers diagrams with exchanges between two legs $i$ and $j$, one finds that they contain cusp logarithms, while diagrams where the gluons attach to a single leg are scaleless for light-like Wilson lines and vanish. In $d$ dimensions, the one-leg contribution can be obtained by the substitutions $n_j\to n_i$ and $\bm{T}_j\to\bm{T}_i$. However, after expanding around $d=4$ the limit $n_j\to n_i$ is singular for the cusp term. 

\begin{figure}
\begin{center}
\psfrag{a}[bl]{$i$}
\psfrag{b}[bl]{$j$}
\psfrag{c}[bl]{$i$}
\includegraphics[height=3.2cm]{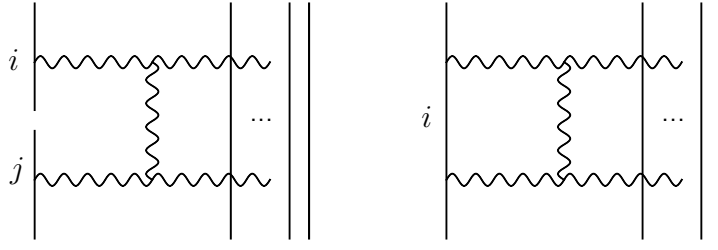}
\end{center}
\vspace{-4mm}
\caption{\label{fig:cancellations}
Graphical illustration of the sum over color-symmetrized soft gluon attachments giving rise to the relations $f_3=4f_6$, $g_1=0$ (left) and $f_2=f_5$ (right).}
\end{figure}

While the naive symmetry argument does not work for the cusp logarithms, we expect it to remain valid for those terms in the amplitude that do not depend on the momentum variables and light-cone vectors of the partons in the set of Wilson lines considered in Figure~\ref{fig:rules}. This assertion should be checked with an explicit calculation. Assuming it is true, we proceed to derive its implications. To this end, consider first the contributions proportional to $f_3$ and $f_6$ in the formula (\ref{Deltags}) for the most general set of extra terms arising at three-loop order. Using relation (\ref{4sym}) and renaming some summation indices, these two terms can be rewritten as
\begin{equation}\label{wonderful}
   \bm{\Delta\Gamma}_s
   \ni f_3(\alpha_s) \sum_{(i,j,k)} \beta_{ij}\,{\cal T}_{ijkk}\,
    + 4f_6(\alpha_s)\!\sum_{(i,j,k,l)}\!\beta_{ij}\,{\cal T}_{ijkl} 
   \stackrel{!}{=} f_3(\alpha_s)\, 
    \sum_{(i,j)}\,\beta_{ij} \sum_{k,l\ne i,j} {\cal T}_{ijkl} \,.
\end{equation}
The last relation follows from the structure of color-symmetrized soft gluon attachments shown in Figure~\ref{fig:rules}, when we take into account that the light-cone vectors $n_k$ and $n_l$ of the Wilson lines not involved in the cusp do not enter the value of the loop integral. It is illustrated in the left diagram depicted in Figure~\ref{fig:cancellations}. Relation (\ref{wonderful}) implies that $4f_6(\alpha_s)=f_3(\alpha_s)$. When combined with relation (\ref{rela1}), this result leads to
\begin{equation}\label{barf10}
   \bar f_1(\alpha_s) = f_1(\alpha_s) - 2 f_4(\alpha_s)
   - 4 f_6(\alpha_s) = 0 \,.
\end{equation}
Consider next the contributions proportional to $f_2$ and $f_5$ in (\ref{Deltags}). They both contain two gluons attached to the Wilson line for parton $i$, plus a sum over the possible attachments of the remaining two gluons. In analogy with (\ref{wonderful}), we conclude that
\begin{equation}
   \bm{\Delta\Gamma}_s
   \ni f_2(\alpha_s)\,\sum_{(i,j)}\,{\cal T}_{iijj}\,
    + f_5(\alpha_s) \sum_{(i,j,k)} {\cal T}_{iijk}
   \stackrel{!}{=} f_2(\alpha_s)\,\sum_i 
   \sum_{j,k\ne i} {\cal T}_{iijk} \,.
\end{equation}
This relation, which is illustrated in the right diagram in Figure~\ref{fig:cancellations}, implies that
\begin{equation}\label{barf20}
   \bar f_2(\alpha_s) = f_2(\alpha_s) - f_5(\alpha_s) = 0 \,.
\end{equation}
We conclude that the contributions of $\bar f_1(\alpha_s)$ and $\bar f_2(\alpha_s)$ to $\bm{\Delta\Gamma}_3$ in (\ref{DG3final}) vanish, which leaves only the possibility of a contribution involving the function $F(x,y)$ of conformal cross ratios. The physical argument underlying this cancellation is the vanishing overall color charge of the $n$-parton scattering amplitude, combined with the simple form of color-symmetrized soft gluon attachments to collinear particles.

For the new structure at four-loop order involving higher Casimir invariants a similar argument can be made. In analogy with (\ref{wonderful}), the three structures shown in (\ref{eq:hc1}) should all derive from
\begin{equation}
   \sum_{(i,j)}\,\beta_{ij} \sum_{k,l\ne i,j} {\cal D}_{ijkl} 
   = \sum_{(i,j)}\,\beta_{ij}\,\bigg[ 
    2 {\cal D}_{iijj} + 2 {\cal D}_{iiij} 
    - \sum_{k\ne i,j}\,{\cal D}_{ijkk} \bigg] \,.
\end{equation}
It follows that $g_1(\alpha_s) = g_2(\alpha_s) = - 2 g_3(\alpha_s)$, which is incompatible with (\ref{g3rela}) unless we require that
\begin{equation}\label{g10}
   g_1(\alpha_s) = 0 \,,
\end{equation}
so that $\bm{\Delta\Gamma}_4^{\rm cusp}$ in (\ref{DG4cusp}) vanishes. Hence, we find that the cusp anomalous dimension obeys Casimir scaling also at four-loop order. This observation is not in contradiction to the fact that, on physical grounds, one expects the finite terms in the vacuum expectation values of Wilson loops to receive contributions from higher Casimir invariants \cite{Frenkel:1984pz}.

\subsection{Two-parton collinear limits}
\label{sec:collinearlimit}

We will now rederive the conditions (\ref{barf10}), (\ref{barf20}), and (\ref{g10}) from an independent consideration. To this end, we analyze the behavior of the extra terms in $\bm{\Delta\Gamma}_3$ given in (\ref{DG3final}) in the two-particle collinear limit and check whether they are compatible with collinear factorization. For the contributions of the first two new structures to the anomalous dimension of the splitting amplitudes, we obtain
\begin{equation}
\begin{split}
   \bm{\Delta\Gamma}_{\rm Sp}(\{p_1,p_2\},\mu)
    \big|_{\bar f_1(\alpha_s)}
   &= 2\!\sum_{(i,j)\ne 1,2}\!{\cal T}_{12ij} \left[ 
    \ln\frac{(-s_{Pi})(-s_{Pj})}{(-s_{12})(-s_{ij})} 
    + \ln z(1-z) \right] , \\[-0.05mm]
   \bm{\Delta\Gamma}_{\rm Sp}(\{p_1,p_2\},\mu)
    \big|_{\bar f_2(\alpha_s)}
   &= 2 {\cal T}_{1122} - 4\,\sum_{i\ne 1,2}\,{\cal T}_{12ii} \,.
\end{split}
\end{equation}
Both contributions are incompatible with the factorization of collinear singularities, because the splitting  amplitudes and their anomalous dimension must not depend on the colors and momenta of the remaining partons not involved in the splitting process. We must therefore require that $\bar f_1(\alpha_s)=\bar f_2(\alpha_s)=0$, in accordance with (\ref{barf10}) and (\ref{barf20}). 

An analogous calculation shows that the new structure proportional to the function $g_1(\alpha_s)$ in the expression for $\bm{\Delta\Gamma}_4^{\rm cusp}$ in (\ref{DG4cusp}), which would lead to a violation of Casimir scaling of the cusp anomalous dimension, is incompatible with the two-particle collinear limits. Considering the terms proportional to $\ln[\mu^2/(-s_{12})]$ for example, we find that
\begin{equation}
   \bm{\Delta\Gamma}_{\rm Sp}(\{p_1,p_2\},\mu)
    \big|_{g_1(\alpha_s)}
   = - 2\,\bigg[ {\cal D}_{1122} + {\cal D}_{1112}
    + {\cal D}_{1222} + \sum_{i\ne 1,2} {\cal D}_{12ii} \bigg]\,
    \ln\frac{\mu^2}{-s_{12}} + \dots \,.
\end{equation}
The sum over ${\cal D}_{12ii}$ color structures cannot be expressed in terms of the color generators of partons 1 and 2 alone. Hence, we must require that $g_1(\alpha_s)=0$, in agreement with (\ref{g10}).

Let us then finally consider the third structure in (\ref{DG3final}). The conformal cross ratios either vanish or diverge when two parton momenta become collinear. In order to study the collinear limits properly, we adopt the following parameterization of the momenta of partons 1 and 2:
\begin{equation}
   p_1^\mu = zE n^\mu + p_\perp^\mu 
    - \frac{p_\perp^2}{4zE}\,\bar n^\mu \,, \qquad
   p_2^\mu = (1-z)E n^\mu - p_\perp^\mu 
    - \frac{p_\perp^2}{4(1-z)E}\,\bar n^\mu \,, 
\end{equation}
where $n^2=\bar n^2=0$ and $n\cdot\bar n=2$, and the ratio $p_\perp/E$ is a small expansion parameter. The collinear limits corresponds to taking $p_\perp\to 0$ at fixed $E$. This parameterization is such that $p_1^2=p_2^2=0$ remain on-shell, while $-s_{12}=p_\perp^2/[z(1-z)]$. The contribution to the anomalous dimension of the splitting amplitudes resulting from the last term in (\ref{DG3final}) can then be written as
\begin{equation}\label{f7contr}
   \bm{\Delta\Gamma}_{\rm Sp}(\{p_1,p_2\},\mu) \big|_F
   = \!\sum_{(i,j)\ne 1,2}\!\Big[ 
   8 {\cal T}_{12ij}\,F(\omega_{ij},\omega_{ij})
   + 4 {\cal T}_{1ij2}\,F(\epsilon_{ij},-2\omega_{ij}) \Big] \,,
\end{equation}
where, at leading power in $p_\perp/E$,
\begin{equation}
\begin{split}
   \epsilon_{ij}\equiv\beta_{1ij2}
   &= \frac{1}{z(1-z) E} 
    \left( \frac{p_\perp\cdot p_i}{n\cdot p_i}
    - \frac{p_\perp\cdot p_j}{n\cdot p_j} \right) \to 0 \,, \\
   \omega_{ij}\equiv\beta_{12ij}
   &= \ln\frac{p_\perp^2}{4z^2(1-z)^2E^2}
    + \ln\frac{(-s_{ij})}{(-n\cdot p_i)(-n\cdot p_j)}
   \to - \infty \,.
\end{split}
\end{equation} 
Each of the two terms on the right-hand side of (\ref{f7contr}) is incompatible with the collinear factorization constraint (\ref{Gsplit}), unless the coefficient functions vanish in the collinear limit $p_\perp\to 0$.\footnote{We are grateful to Lance Dixon for pointing out that a contribution to $\bm{\Delta\Gamma}_3$ in (\ref{DG3final}) involving a function of conformal cross ratios that vanishes in all collinear limits in not excluded by our analysis.}
Since the splitting amplitudes scale like $1/\sqrt{s_{12}}\sim 1/|p_\perp|$, the functions must fall off at least as fast as $\epsilon_{ij}\sim e^{\omega_{ij}/2}$. It is an open question whether a function of transcendentality up to 5 (since it appears in the $1/\epsilon$ pole term at three-loop order) with these properties exists and appears in the soft anomalous-dimension matrix at three-loop order. The validity of our conjecture rests on the assumption that $F(x,y)=0$.

We note in this context that a conjecture about the exponentiation of the {\em finite\/} terms of scattering amplitudes in ${\cal N}=4$ SYM theory \cite{Bern:2005iz} was recently shown to be invalidated, for the case of $n>5$ partons and at two-loop order, by Regge cut contributions \cite{Bartels:2008ce}, which vanish in all two-parton collinear limits \cite{Bern:2008ap}. However, these contributions were found not to affect the divergent terms of the amplitude \cite{Bartels:2008sc}.

\subsection{Extension to higher orders}

Leaving aside the possibility of functions of conformal cross ratios that vanish in all collinear limits, the arguments presented in the previous sections establish our conjecture (\ref{magic}) at three-loop order and moreover exclude a certain class of modifications at four-loop order. It would certainly be worthwhile to test the rigor of these arguments with explicit multi-loop calculations, as we have emphasized toward the end of Section~\ref{sec:simplifications}. Nevertheless, in our opinion these arguments provide compelling evidence that our result is correct to all (finite) orders in perturbation theory. Essentially, the constraint (\ref{ourconstraint}) derived from the factorization properties of SCET, when combined with the splitting relation (\ref{Gsplit}), requires that the anomalous-dimension matrix must be {\em linear\/} in both the cusp angles and the color generators of the external partons, and that the coefficient of the cusp term is the cusp anomalous dimension. This implies that momentum-independent terms are color-diagonal to all orders. Momentum-dependent structures must have the color-dipole structure exhibited in (\ref{magic}).

It thus appears that our relation (\ref{magic}) may indeed be an exact result of perturbative quantum field theory, valid in arbitrary massless gauge theories. There are few such results known in the literature, and it is not unreasonable to expect that the discovery of this relation will have profound implications for our understanding of scattering amplitudes.

\section{Summary and outlook}
\label{sec:summary}

We have shown that the IR poles of on-shell scattering amplitudes in massless QCD can be mapped onto the UV poles of the renormalization factor $\bm{Z}$ of $n$-jet operators in SCET. The RG evolution of these operators is governed by a universal anomalous-dimension matrix, whose form is severely constrained by soft-collinear factorization, non-abelian exponentiation, and the behavior of amplitudes in collinear limits.  We have argued that only the simple form
\[
   \bm{\Gamma}(\{\underline{p}\},\mu) 
   = \sum_{(i,j)}\,\frac{\bm{T}_i\cdot\bm{T}_j}{2}\,
   \gamma_{\rm cusp}(\alpha_s)\,\ln\frac{\mu^2}{-s_{ij}} 
   + \sum_i\,\gamma^i(\alpha_s)
\]
is consistent with all these constraints and have explicitly checked that they exclude any additional contributions up to three-loop accuracy. We also find that contributions from terms involving higher Casimir operators are excluded at four loops. However, our arguments do not exclude the presence of the term
\[
   \Delta\bm{\Gamma}(\{\underline{p}\},\mu) 
   =  \sum_{(i,j,k,l)}\!f^{ade} f^{bce}\,
   \bm{T}_i^a\,\bm{T}_j^b\,\bm{T}_k^c\,\bm{T}_l^d\,
   F(\beta_{ijkl},\beta_{iklj}-\beta_{iljk})
\]
at three-loop order (and analogous terms in higher orders), where the function $F(x,y)$ would have to vanish whenever two parton momenta become collinear, and the conformal cross ratios $\beta_{ijkl}$ are defined in (\ref{CCR}). We consider it unlikely that such functions arise in the anomalous-dimension matrix and thus conjecture that they are absent. Since the discussion in our paper relies solely on the commutation relations and the Jacobi identity, our results apply to any massless gauge theory based on a semi-simple group. Furthermore, by combining our results with methods developed in \cite{Catani:2000ef,Glover:2001ev,Penin:2005kf,Mitov:2006xs,Becher:2007cu}, which relate the singularities of massive and massless amplitudes, our formalism can be generalized to the massive case. This is worked out in detail in \cite{Becher:2009kw}.

The above form of the anomalous dimension is consistent with all existing results for higher-order scattering amplitudes, but it would be desirable to further test it with explicit multi-loop calculations. It will be particularly interesting to compare with the three-loop result for the full four-parton amplitude in ${\cal N}=4$ SYM given in \cite{Bern:2008pv}, once the necessary master integrals become available. In particular, this result will check whether color correlations between four partons appear, and whether they obey the constraints from soft-collinear factorization and collinear limits, i.e.\ whether they have the form discussed above. Also of great interest would be a calculation of the four-loop cusp anomalous dimensions of quarks and gluons in QCD or its supersymmetric extensions, either by direct calculation, extending the work of \cite{Moch:2004pa}, or by using the approach based on the AdS/CFT correspondence \cite{Alday:2007mf}. This would test our prediction of Casimir scaling. The recent accomplishment of the exact evaluation of three-loop form factor integrals \cite{Baikov:2009bg, Heinrich:2009be} gives us hope that these calculations will become feasible in the not too distant future. 

Understanding the IR structure of scattering amplitudes is of significant theoretical interest, and having explicit results for the divergent part of the amplitudes provides an important check on multi-loop calculations. Also, since the singularities must cancel against those of diagrams with real gluon emission, our results might lead to an improved treatment of the soft and collinear singularities in real emission processes. However, the most important application of our work are resummations of Sudakov logarithms in multi-jet processes. 
There is a rich literature on Sudakov resummation for QCD processes, starting with the pioneering papers \cite{Sterman:1986aj,Catani:1989ne,Magnea:1990qg,Korchemsky:1993uz} (for a review, see \cite{Kidonakis:1999ze} and references therein). In the effective theory, the resummation of these logarithmically-enhanced contributions is achieved by solving the RG equations for the Wilson coefficients. For two-jet observables, effective-theory methods have been used to perform resummations to N$^3$LL accuracy. Examples include threshold resummation for deep-inelastic scattering \cite{Becher:2006mr}, Drell-Yan production \cite{Idilbi:2006dg,Becher:2007ty}, Higgs-boson production \cite{Idilbi:2005ni,Ahrens:2008nc}, and the extraction of $\alpha_s$ from $e^+ e^-\to\mbox{2 jets}$ \cite{Becher:2008cf}. With the anomalous dimension for the $n$-jet case at hand, it now becomes possible to reach the same accuracy also for more complicated observables. The evolution equation (\ref{RGE}) is simple enough to admit exact solutions for a given $n$-parton scattering process. This can be used to perform the resummation of large Sudakov logarithms in closed form. 

A lot of work has been done to match parton showers with fixed-order calculations. The effective-theory approach allows one to not only combine fixed-order with leading-log resummations, but also to systematically resum subleading Sudakov logarithms. To obtain predictions for $n$-jet observables, one needs the fixed-order results for $n$-parton amplitudes, which correspond to the Wilson coefficients of the operators in SCET. By combining our results with a tree-level matrix element generator, one can obtain NLL resummations. New efficient methods for one-loop calculations of processes with many legs are currently used to develop generators for one-loop matrix elements \cite{Bern:2008ef}. Solving the associated RG equations then leads to NNLL predictions. Finally, for $2\to2$ processes, the virtual corrections are known to two-loop accuracy, so that in this case N$^3$LL accuracy can  be achieved. The predictions for production rates of $n$-jet processes are obtained by combining the resummed hard-scattering Wilson coefficients with jet and soft functions. In contrast to fixed-order calculations or parton showers, these predictions are inclusive in the sense that they predict jet observables, and not the contributions of individual partons: the jet functions already include the integration over the phase space of the partons within a jet. 

For the analysis of this paper the concepts of effective field theory, such as mode separation and RG methods, were helpful to address an old problem of perturbative quantum field theory, which had resisted a solution using traditional methods. Effective field theory methods provide a natural language to discuss multi-scale problems, and we believe that these methods will play an important role in improving the accuracy of predictions for collider processes.

\subsection*{Acknowledgments}

We are grateful to Nima Arkani-Hamed, Adi Armoni, Bill Bardeen, Lance Dixon, Keith Ellis, Walter Giele, Uli Haisch, Juan Maldacena, and Alexander Penin for stimulating discussions and exchanges, and to Dominik Stoeckinger for pointing out a typo in eq.~(17) of an earlier version of this paper. One of us (M.N.) thanks Christian Bauer for discussions of related questions as early as in the spring of 2007. The research of T.B.\ was supported by the U.S.\ Department of Energy under Grant DE-AC02-76CH03000. Fermilab is operated by the Fermi Research Alliance under contract with the Department of Energy.

\newpage
\begin{appendix}
\renewcommand{\theequation}{A\arabic{equation}}
\setcounter{equation}{0}

\section{Three-loop anomalous dimensions and $\bm{Z}$-factor}

The three-loop expression for the renormalization factor $\bm{Z}$ removing the IR poles of $n$-parton scattering amplitudes, as shown in (\ref{renorm}), can be obtained by exponentiating our result (\ref{reslnZ}). An alternative way to determine the $\bm{Z}$-matrix is to use the relations \cite{Becher:2005pd}
\begin{equation}
\begin{split}
   \bm{\Gamma} 
   &= 2\alpha_s\,\frac{\partial}{\partial\alpha_s}\,\bm{Z}^{(1)}
    \,, \\
   2\alpha_s\,\frac{\partial\bm{Z}^{(n+1)}}{\partial\alpha_s} 
   &= \bm{\Gamma}\,\bm{Z}^{(n)} 
    + \beta(\alpha_s)\,\frac{\partial\bm{Z}^{(n)}}{\partial\alpha_s}
    + \frac{\partial\bm{Z}^{(n)}}{\partial\ln\mu} \,,
\end{split}
\end{equation}
where the superscript ``$(n)$'' denotes the coefficient of the $1/\epsilon^n$ pole term. These relations allow one to construct the higher $1/\epsilon^n$ pole terms in a recursive way. Either way, we find
\begin{eqnarray}\label{result}
   \bm{Z} &=& 1 + \frac{\alpha_s}{4\pi} \!
    \left( \frac{\Gamma_0'}{4\epsilon^2}
    + \frac{\bm{\Gamma}_0}{2\epsilon} \right) 
   + \left( \frac{\alpha_s}{4\pi} \right)^2 \! \left[
    \frac{(\Gamma_0')^2}{32\epsilon^4} 
    + \frac{\Gamma_0'}{8\epsilon^3} 
    \left( \bm{\Gamma}_0 - \frac32\,\beta_0 \right) 
    + \frac{\bm{\Gamma}_0}{8\epsilon^2} 
    \left( \bm{\Gamma}_0 -2\beta_0 \right) 
    + \frac{\Gamma_1'}{16\epsilon^2}
    + \frac{\bm{\Gamma}_1}{4\epsilon} \right] \nonumber\\
   &&\mbox{}+ \left( \frac{\alpha_s}{4\pi} \right)^3 \Bigg[
    \frac{(\Gamma_0')^3}{384\epsilon^6}
    + \frac{(\Gamma_0')^2}{64\epsilon^5}
     \left( \bm{\Gamma}_0 - 3\beta_0 \right)
    + \frac{\Gamma_0'}{32\epsilon^4}
     \left( \bm{\Gamma}_0 - \frac43\,\beta_0 \right)
     \left( \bm{\Gamma}_0 - \frac{11}{3}\,\beta_0 \right)
    + \frac{\Gamma_0'\Gamma_1'}{64\epsilon^4} \nonumber\\
   &&\hspace{1.9cm}\mbox{}+ \frac{\bm{\Gamma}_0}{48\epsilon^3}
     \left( \bm{\Gamma}_0 - 2\beta_0 \right)
     \left( \bm{\Gamma}_0 - 4\beta_0 \right)
    + \frac{\Gamma_0'}{16\epsilon^3}
     \left( \bm{\Gamma}_1 - \frac{16}{9}\,\beta_1 \right)
    + \frac{\Gamma_1'}{32\epsilon^3}
     \left( \bm{\Gamma}_0 - \frac{20}{9}\,\beta_0 \right) \nonumber\\
   &&\hspace{1.9cm}\mbox{}+ 
    \frac{\bm{\Gamma}_0\bm{\Gamma}_1}{8\epsilon^2} 
    - \frac{\beta_0\bm{\Gamma}_1+\beta_1\bm{\Gamma}_0}{6\epsilon^2}
    + \frac{\Gamma_2'}{36\epsilon^2}
    + \frac{\bm{\Gamma}_2}{6\epsilon} \Bigg] 
    + {\cal O}(\alpha_s^4) \,.
\end{eqnarray}
The expansion coefficients of the anomalous-dimensions and $\beta$-function have been defined in (\ref{Gbexp}). Through relations (\ref{magic}) and (\ref{Gampr}), the coefficients $\bm{\Gamma}_n$ and $\Gamma_n'$ can in turn be expressed in terms of the expansion coefficients of the anomalous dimensions $\gamma_{\rm cusp}$, $\gamma^q$, and $\gamma^g$, defined in analogy with the first relation in (\ref{Gbexp}). 

We now list these coefficients up to three-loop order in the $\overline{{\rm MS}}$ renormalization scheme. The expansion of the cusp anomalous dimension $\gamma_{\rm cusp}$ to two-loop order was obtained some time ago \cite{Korchemsky:1987wg,Korchemsky:1988hd,Korchemskaya:1992je,Kodaira:1981nh,Catani:1988vd}. The three-loop coefficient was calculated in \cite{Moch:2004pa}. The results are
\begin{eqnarray}
   \gamma_0^{\rm cusp} &=& 4 \,, \nonumber\\
   \gamma_1^{\rm cusp} &=& \left( \frac{268}{9} 
    - \frac{4\pi^2}{3} \right) C_A - \frac{80}{9}\,T_F n_f \,,
    \nonumber\\
   \gamma_2^{\rm cusp} &=& C_A^2 \left( \frac{490}{3} 
    - \frac{536\pi^2}{27}
    + \frac{44\pi^4}{45} + \frac{88}{3}\,\zeta_3 \right) 
    + C_A T_F n_f  \left( - \frac{1672}{27} + \frac{160\pi^2}{27}
    - \frac{224}{3}\,\zeta_3 \right) \nonumber\\
   &&\mbox{}+ C_F T_F n_f \left( - \frac{220}{3} + 64\zeta_3 \right) 
    - \frac{64}{27}\,T_F^2 n_f^2 \,.
\end{eqnarray}
The anomalous dimension $\gamma^q=\gamma^{\bar q}$ can be determined from the three-loop expression for the divergent part of the on-shell quark form factor in QCD \cite{Moch:2005id}. The result was extracted in \cite{Becher:2006mr}. In the notation of this paper $2\gamma^q=\gamma^V$. We obtain
\begin{eqnarray}
   \gamma_0^q &=& -3 C_F \,, \nonumber\\
   \gamma_1^q &=& C_F^2 \left( -\frac{3}{2} + 2\pi^2
    - 24\zeta_3 \right)
    + C_F C_A \left( - \frac{961}{54} - \frac{11\pi^2}{6} 
    + 26\zeta_3 \right)
    + C_F T_F n_f \left( \frac{130}{27} + \frac{2\pi^2}{3} \right) ,
    \nonumber\\
   \gamma_2^q &=& C_F^3 \left( -\frac{29}{2} - 3\pi^2
    - \frac{8\pi^4}{5}
    - 68\zeta_3 + \frac{16\pi^2}{3}\,\zeta_3 + 240\zeta_5 \right) 
    \nonumber\\
   &&\mbox{}+ C_F^2 C_A \left( - \frac{151}{4} + \frac{205\pi^2}{9}
    + \frac{247\pi^4}{135} - \frac{844}{3}\,\zeta_3
    - \frac{8\pi^2}{3}\,\zeta_3 - 120\zeta_5 \right) \nonumber\\
   &&\mbox{}+ C_F C_A^2 \left( - \frac{139345}{2916} - \frac{7163\pi^2}{486}
    - \frac{83\pi^4}{90} + \frac{3526}{9}\,\zeta_3
    - \frac{44\pi^2}{9}\,\zeta_3 - 136\zeta_5 \right) \nonumber\\
   &&\mbox{}+ C_F^2 T_F n_f \left( \frac{2953}{27} - \frac{26\pi^2}{9} 
    - \frac{28\pi^4}{27} + \frac{512}{9}\,\zeta_3 \right) 
    \nonumber\\
   &&\mbox{}+ C_F C_A T_F n_f \left( - \frac{17318}{729}
    + \frac{2594\pi^2}{243} + \frac{22\pi^4}{45} 
    - \frac{1928}{27}\,\zeta_3 \right) \nonumber\\
   &&\mbox{}+ C_F T_F^2 n_f^2 \left( \frac{9668}{729} 
    - \frac{40\pi^2}{27} - \frac{32}{27}\,\zeta_3 \right) .
\end{eqnarray}
Similarly, the expression for the gluon anomalous dimension can be extracted from the divergent part of the gluon form factor obtained in \cite{Moch:2005id}. In terms of the anomalous dimensions given in \cite{Becher:2007ty}, we have $2\gamma^g(\alpha_s)=\gamma^t(\alpha_s)+\gamma^S(\alpha_s)+\beta(\alpha_s)/\alpha_s$. We find
\begin{eqnarray}
   \gamma_0^g &=& - \beta_0 
    = - \frac{11}{3}\,C_A + \frac43\,T_F n_f \,, \nonumber\\
   \gamma_1^g &=& C_A^2 \left( -\frac{692}{27} + \frac{11\pi^2}{18}
    + 2\zeta_3 \right) 
    + C_A T_F n_f \left( \frac{256}{27} - \frac{2\pi^2}{9} \right)
    + 4 C_F T_F n_f \,, \nonumber\\
   \gamma_2^g &=& C_A^3 \left( - \frac{97186}{729} 
    + \frac{6109\pi^2}{486} - \frac{319\pi^4}{270} 
    + \frac{122}{3}\,\zeta_3 - \frac{20\pi^2}{9}\,\zeta_3 
    - 16\zeta_5 \right) \nonumber\\
   &&\mbox{}+ C_A^2 T_F n_f \left( \frac{30715}{729}
    - \frac{1198\pi^2}{243} + \frac{82\pi^4}{135} 
    + \frac{712}{27}\,\zeta_3 \right) \nonumber\\
   &&\mbox{}+ C_A C_F T_F n_f \left( \frac{2434}{27} 
    - \frac{2\pi^2}{3} - \frac{8\pi^4}{45} 
    - \frac{304}{9}\,\zeta_3 \right) 
    - 2 C_F^2 T_F n_f \nonumber\\
   &&\mbox{}+ C_A T_F^2 n_f^2 \left( - \frac{538}{729}
    + \frac{40\pi^2}{81} - \frac{224}{27}\,\zeta_3 \right) 
    - \frac{44}{9}\,C_F T_F^2 n_f^2 \,.
\end{eqnarray}
Our results for $\gamma^q$ and $\gamma^g$ are valid in conventional dimensional regularization, where polarization vectors and spinors of all particles are treated as $d$-dimensional objects (so that gluons have $(2-2\epsilon)$ helicity states). In the 't~Hooft-Veltman scheme \cite{'tHooft:1972fi} or the four-dimensional helicity scheme \cite{Bern:1991aq}, their values would be different.

\renewcommand{\theequation}{B\arabic{equation}}
\setcounter{equation}{0}

\section{Leading-color limit\label{sec:leadingcolor}}

Let us briefly discuss the $N_c\to\infty $ limit, where a number of three- and four-loop results are available for ${\cal N}=4$ SYM. It is interesting to ask whether these provide a test of our conjecture for the anomalous dimension. Unfortunately, it turns out that both the terms in (\ref{DG3final}) as well as the higher Casimir terms in  (\ref{DG4cusp}) are subleading in the large-$N_c$ limit and are thus not constrained by known results for planar amplitudes.

A basis of leading color structures of $n$-particle gluonic amplitudes is given by the traces $\mbox{tr}(t^{a_1}\ldots\,t^{a_n})$ of color matrices in the fundamental representation of the gauge group \cite{Bern:1991aq,Berends:1987cv,Mangano:1987xk} (see \cite{Dixon:1996wi} for a pedagogical review). The cyclicity of the trace implies that there are $(n-1)!$ different color structures for $n$ gluons. Structures involving several traces are subleading for $N_c\to\infty$. These color structures can be viewed as the different possibilities of attaching $n$ gluons to a quark loop. The leading contributions to squared amplitudes in the $N_c\to\infty$ limit arise when a given color structure is contracted with its reverse. This gives
\begin{equation}
   \mbox{tr}\left( t^{a_1}\ldots\,t^{a_n} \right)\, 
   \mbox{tr}\left( t^{a_n}\ldots\,t^{a_1} \right) 
   = \frac{N_c^n}{2^n} \,,
\end{equation}
where terms of subleading order have been dropped. All other contractions are subleading, so in this sense the color basis is orthogonal.

For $N_c\to\infty$, the color structure $ {\bm T}_i\cdot {\bm T}_j$ acts on these traces in a particularly simple way. For $n>2$, we find
\begin{equation}
\begin{split}
   (\bm{T}_i)^{a_i b_i}\cdot(\bm{T}_{i+1})^{a_{i+1} b_{i+1}}\,
   &\mbox{tr}\left( t^{a_1}\ldots\,t^{b_i}\,t^{b_{i+1}}\ldots\,t^{a_n} 
    \right) \\
   &\!\!= (-i) f^{a_i b_i c}\,(-i) f^{a_{i+1} b_{i+1} c}\,
    \mbox{tr}\left( t^{a_1}\ldots\,t^{b_i}\,t^{b_{i+1}}\ldots\,t^{a_n} 
    \right) \\
   &\!\!= - \frac{N_c}{2}\,\mbox{tr}\left( t^{a_1}\ldots\,t^{a_i}\,
    t^{a_{i+1}}\ldots\,t^{a_n} \right) ,  
\end{split}
\end{equation}
where again subleading terms have been dropped. Note that the result is leading order in color, since in the anomalous dimension this color structure gets multiplied by $g^2$, and $g^2 N_c$ is held fixed in the $N_c\to\infty$ limit. For the case of ${\bm T}_i\cdot {\bm T}_j$ with $i$ and $j$ not adjacent we find a result that is of subleading order. In operator notation, we thus obtain
\begin{equation}
   {\bm T}_i\cdot{\bm T}_j \to - \frac{N_c}{2}\,\delta_{j,i\pm 1} \,.
\end{equation}
For the additional color structures arising in (\ref{DG3final}), we find
\begin{equation}\label{eq:extracolor}
\begin{split}
   f^{ade} f^{bce}\,({\bm T}_i^a\,{\bm T}_i^b)_+\,
    {\bm T}_{j}^c\,{\bm T}_{k}^d 
   &= {\cal O}(N_c) \,, \\
   f^{ade} f^{bce}\,{\bm T}_i^a\,{\bm T}_j^b\,
    {\bm T}_k^c\,{\bm T}_l^d 
   &= {\cal O}(N_c) \,,
\end{split}
\end{equation}
for all indices $i,j,k,l$ different. Similarly, for the new structures in (\ref{DG4cusp}) we obtain
\begin{equation}\label{eq:extracolor2}
   {\cal D}_{iijj} = {\cal O}(N_c) \,, \qquad
   {\cal D}_{iiij} = {\cal O}(N_c^3) \,, \qquad
   {\cal D}_{ijkk} = {\cal O}(N_c) \,.
\end{equation}
At $n$-loop order the leading color structures are of the form $N_c^n\,\mbox{tr}(t^{a_1}\ldots\,t^{a_n})$. Since the two structures (\ref{eq:extracolor}) appear first at three-loop order, while those in (\ref{eq:extracolor2}) arise first at the level of four loops, their contribution is suppressed compared to the leading term.

\end{appendix}

\end{document}